\documentclass[%
twocolumn,
superscriptaddress,
 amsmath,amssymb,
 aps,
pra,
longbibliography,
floatfix,
]{revtex4-2}

\usepackage{graphicx}
\usepackage{dcolumn}
\usepackage{bm}
\usepackage{aas_macros}
\usepackage[acronym]{glossaries}
\usepackage[tight]{units}
\usepackage{xspace}
\usepackage{hyperref}
\usepackage[svgnames]{xcolor}
\usepackage{soul}
\usepackage[normalem]{ulem}

\newacronym{SRMHD}{SRMHD}{special-relativistic magnetohydrodynamics}
\newacronym{EOS}{EOS}{equation of state}
\newacronym{GW}{GW}{gravitational wave}
\newacronym{NSE}{NSE}{nuclear statistical equilibrium}
\newacronym[shortplural={CCSNe}, longplural={core collapse supernovae}]{CCSN}{CCSN}{core collapse supernova}
\newacronym{PNS}{PNS}{proto-neutron star}
\newacronym{SASI}{SASI}{stationary accretion shock instability}
\newacronym{1D}{1D}{1-dimensional}
\newacronym{2D}{2D}{2-dimensional}
\newacronym{3D}{3D}{3-dimensional}
\newacronym{BNS}{BNS}{binary neutron star}
\newacronym{ET}{ET}{Einstein Telescope}
\newacronym{CE}{CE}{Cosmic Explorer}
\newacronym{ASD}{ASD}{amplitude spectral density}
\newacronym{FTA}{FTA}{Fourier transform amplitude}
\newacronym{SNR}{SNR}{signal-to-noise ratio}

\received{}

\accepted{}

\begin{document}

\preprint{APS/123-QED}

\title{Resonant amplification of multimessenger emission in rotating stellar core collapse}

\author{Marco~Cusinato}\email{marco.cusinato@uv.es}
\affiliation{%
 Departament d'Astronomia i Astrof\'{\i}sica, Universitat de Val\`encia, Av.~Vicent Andrés Estellés 19, 46100, Burjassot (Val\`encia), Spain
} 

\author{Martin~Obergaulinger}
\email{martin.obergaulinger@uv.es}
\affiliation{%
 Departament d'Astronomia i Astrof\'{\i}sica, Universitat de Val\`encia, Av.~Vicent Andrés Estellés 19, 46100, Burjassot (Val\`encia), Spain
}
\affiliation{Observatori Astronòmic, Universitat de València, 46980 Paterna, Spain}
\author{Miguel~Á.~Aloy}
\email{miguel.a.aloy@uv.es}
\affiliation{%
 Departament d'Astronomia i Astrof\'{\i}sica, Universitat de Val\`encia, Av.~Vicent Andrés Estellés 19, 46100, Burjassot (Val\`encia), Spain
} 
\affiliation{Observatori Astronòmic, Universitat de València, 46980 Paterna, Spain}
\author{Jos\'e~A.~Font}
\affiliation{%
 Departament d'Astronomia i Astrof\'{\i}sica, Universitat de Val\`encia, Av.~Vicent Andrés Estellés 19, 46100, Burjassot (Val\`encia), Spain
}
\affiliation{Observatori Astronòmic, Universitat de València, 46980 Paterna, Spain}

\date{\today}

\begin{abstract}
In a series of axisymmetric core-collapse supernova simulations extending up to $\sim \unit[2]{s}$, we identify a regime of pre-collapse central rotation rates ($\sim \unit[1]{Hz}$) that greatly enhances the emission of gravitational waves (GWs) during extended periods of time after bounce. The enhancement is a consequence of the resonance between the frequency of the fundamental quadrupolar $^2f$-mode of oscillation of the proto-neutron star and the frequency of the epicyclic oscillations at the boundary of the inner core. We observe periods of about several hundred milliseconds each where the resonance is active. The GW emission enhancement produces a correlated resonant modulation of the associated neutrino signal at the same frequencies. With GW frequencies of $\mathcal{O}(\unit[1]{kHz})$ and strain amplitudes within the sensitivity curves of current and next-generation interferometers at distances of $\mathcal{O}(\unit[1]{Mpc})$, this resonant-amplification mechanism may represent a potential game-changer for unveiling the supernova explosion mechanism through multimessenger astronomy.
\end{abstract}

\maketitle

\section{Introduction}
\label{sec:Intro}
\Glspl{CCSN} mark the final stage in the evolution of massive stars ($M_\textnormal{ZAMS} \gtrsim \unit[8]{M_\odot}$) and are among the prime sources of multimessenger emission as they release \glspl{GW}, neutrinos, and electromagnetic signals. The detection of 25 neutrinos from Supernova 1987A \cite{Bionta87,Hirata87,Alexeyev88}, along with its electromagnetic counterpart \cite{Dotani87,Matz88}, demonstrated that multimessenger observations of \glspl{CCSN} are feasible.

Modern neutrino observatories, such as Super-Kamiokande \cite{Scholberg12}, IceCube \cite{IceCube}, and KM3NeT \cite{KM3NeT}, are expected to detect tens of thousands of neutrinos from a Galactic \gls{CCSN} \cite{Suwa22} or even the collective contribution of cosmological supernovae to the diffuse supernova neutrino background \cite{Martinez-Mirave_2024PhRvD.110j3029}. On the \gls{GW} side, advanced detectors like Advanced LIGO \cite{Aasi15}, Advanced Virgo \cite{Acernese15}, and KAGRA \cite{Akutsu19} are sensitive to \glspl{CCSN} within several kiloparsecs. 
Despite dedicated efforts by the LIGO–Virgo–KAGRA collaboration to detect \glspl{CCSN} at distances of up to $\unit[30]{Mpc}$, no signals have been recorded \cite{LVKO3SNsearch23,LVKSN2023ixf2024}. Nonetheless, prospects for detecting extragalactic events remain promising. Future observatories such as \gls{ET} \cite{Maggiore20} and \gls{CE} \cite{Reitze19} are expected to substantially extend the detection range, while emerging insights into processes that enhance the amplitude of the \gls{GW} emission from \glspl{CCSN} may further improve detection chances.

\Glspl{GW} and neutrinos offer complementary insights into \gls{CCSN} dynamics. For instance, neutrinos were crucial in estimating the gravitational binding energy released in Supernova 1987A \cite{Burrows88}. In contrast, \glspl{GW} reveal signatures of multidimensional fluid instabilities--including \gls{PNS} convection \cite{Muller13, Pan18}, \gls{SASI} \cite{CerdaDuran13,TorresForne18,Andresen17}, and prompt convection--emerging during the early stages of the explosion. Moreover, they can constrain the nuclear \gls{EOS} \cite{Malik18,Raaijmakers20} and yield estimates of key \gls{PNS} parameters \cite{TorresForne19b,Bruel23,Rodriguez23}.

The rotation of the progenitor star is pivotal in shaping the waveform of the resulting \gls{GW} signal and, consequently, the detectability of the event \cite{Shibagaki24, Powell2023}. Specifically, when the core of the progenitor star achieves a certain rotational speed, resonances can develop between the \gls{PNS} oscillation modes and the core's rotational frequency. Such resonances have been identified in simplified \gls{2D} models of rotating relativistic stars \cite{Dimmelmeier2006}, and in the collapse of neutron stars to strange quark stars triggered by phase transitions \cite{Abdikamalov09}. Additionally, during the early phases ($\lesssim \unit[100]{ms}$ post-bounce) of \glspl{CCSN} simulations, rotation can excite specific \gls{PNS} modes \cite{Westernacher-Schneider2019}. However, the full evolution of these resonances through the entire operation of the \glspl{CCSN} engine remains to be comprehensively explored. Previous studies indicate that particular pre-collapse rotation rates can notably amplify the \gls{GW} amplitude, introduce new oscillatory modes, and imprint corresponding signatures in the neutrino signal \cite{Westernacher-Schneider2019}.

In this work we delve into the rich tapestry of multimessenger signals emerging from \gls{CCSN} simulations with different initial rotation rates over a span of $\sim 2\,$s. Our findings reveal that when the stellar core rotates at an intermediate rate ($\approx \unit[1]{Hz}$), it triggers resonant modes in the \gls{PNS} that imprint synchronized modulations on the neutrino signal. Remarkably, these resonant  \glspl{GW} generate substantial strains, even surpassing those at bounce, and manifest within mere  hundreds of milliseconds after collapse.
This striking discovery not only deepens our understanding of the dynamic interplay in stellar explosions but also paves the way for novel insights into their multimessenger signatures.

The remainder of this paper is organised as follows: Section \ref{sec:methods} outlines our numerical setup and presents the progenitor model, initial rotation profiles, and nuclear \gls{EOS} used to perform the \gls{CCSN} simulations. In Section \ref{sec:results} we present the main findings of our study. Finally, in Section \ref{sec:conclusion} we discuss our results and draw our conclusions.

\section{Methods}
\label{sec:methods}
We performed \gls{2D} \gls{CCSN} simulations using the  \texttt{Aenus-ALCAR} code \cite{Obergaulinger_2018JPhG...45h4001,Just15,Obergaulinger_2022MNRAS.512.2489}
that couples special relativistic magnetohydrodynamics with a spectral two-moment neutrino transport scheme and incorporates an approximately relativistic gravitational potential \cite{Marek06}. For densities exceeding $\unit[8 \times 10^{7}]{g/cm^3}$, we employed the SFHo  \gls{EOS} \cite{Steiner13}, accounting for photons, electrons, positrons, nucleons, and a full ensemble of light and heavy nuclei in nuclear statistical equilibrium. The \gls{GW} signal was extracted using the quadrupole formula \cite{Obergaulinger_2006A&A...450.1107}.
The simulations were performed in spherical coordinates on grids with 480 logarithmically spaced radial zones and 128 linearly spaced angular zones, while neutrino energies spanning from $\unit[0]{MeV}$ to $\unit[440]{MeV}$ were resolved with 12 logarithmically spaced bins. Each run was evolved for $\unit[2]{s}$, commencing from the pre-collapse phase. In the Appendix~\ref{app:resolution} we assess the dependence of our results with increased grid resolution.

After an extensive survey of rotating stellar models evolved to several hundreds of milliseconds post-collapse, we identified several simulations that presented resonant coupling between \glspl{GW} and rotation. From these, we selected one case in which this feature is particularly pronounced. 
Specifically, it is a Wolf-Rayet star of subsolar metallicity ($Z = \unit[\mbox{0.02}]{Z_\odot}$) and zero-age main sequence mass of $M_\textnormal{ZAMS} = \unit[17]{M_\odot}$, evolved with both rotation and magnetic fields up to the pre-collapse stage \cite{AguileraDena18}. Its initially high rotational velocity, coupled with enhanced rotational mixing, drives chemically homogeneous evolution. Building on this model, and reconstructing the magnetic field topology from the saturation strengths originally included in the (one-dimensional) stellar evolution model (of the order of $10^{9}\,$G inside the iron core), we explored three rotational configurations: the original (\texttt{SR}), with central rotational rate of $\Omega_{\rm c}=\unit[0.29]{rad/s}$, an intermediate case (\texttt{IR}) with a rate amplified by a factor of 3.5, and a fast-rotating scenario (\texttt{FR}) with a 12-fold increase. Although these increased rotation rates are substantial, they remain well within the computed range for massive stars \cite{Woosley_2006ApJ...637..914,Griffiths_2022A&A...665A.147}.

\section{Results}
\label{sec:results}
\begin{figure}
\includegraphics[width=0.5\textwidth]{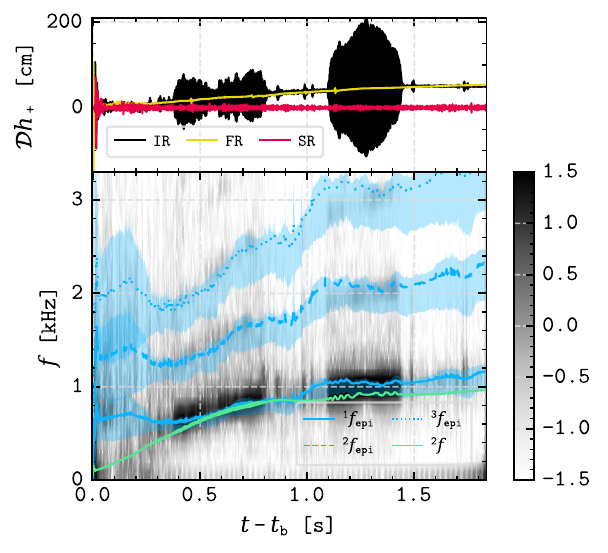} 
\caption{
    \label{fig:GW_spectro}Top panel: \gls{GW} strains emitted by a source at distance $\mathcal{D}$ for models \texttt{SR} (red), \texttt{IR} (black), and \texttt{FR} (yellow). Bottom panel: Spectrogram of the \gls{GW} signal for model \texttt{IR}. Blue lines represent the fundamental frequency (solid), the first (dashed) and the second (dotted) overtones of the epicyclic frequency, with shaded regions showing the associated uncertainties. The green line indicates the fundamental quadrupolar mode, computed using the quasi-universal relation from 
    \cite{TorresForne19}.
}
\end{figure}

The plus polarisation of the \gls{GW} signal, $h_+$, for a source at distance $\mathcal{D}$ (see top panel of Figure~\ref{fig:GW_spectro}) reveals the brief yet intense bounce signal characteristic of rapidly rotating cores \cite{Abdikamalov14,Richers17}. In the most rapidly rotating model \texttt{FR}, the signal amplitude reaches $\mathcal{D}h_+ \gtrsim \unit[100]{cm}$. Following the bounce, the waveform transitions into a regime of weaker oscillations that remain centred around zero in model \texttt{SR} or gradually drift towards positive values in models \texttt{IR} and \texttt{FR}, which ultimately explode via polar jets. 

Unlike models \texttt{SR} and \texttt{FR}, where the oscillatory amplitudes remain relatively constant throughout the simulation, model \texttt{IR} undergoes two distinct phases of pronounced amplitude enhancement, the first occurring between $t \approx \unit[0.4]{s}$ and $\unit[0.8]{s}$, and the second one between $t \approx \unit[1.1]{s}$ and $\unit[1.4]{s}$. During these bursts, the strain range widens to approximately  $\sim \unit[130]{cm}$ and $\sim \unit[300]{cm}$, respectively, reaching levels comparable to the bounce signal observed in model \texttt{FR}.

The bottom panel of Figure~\ref{fig:GW_spectro} shows the \gls{GW} spectrogram of model \texttt{IR}, computed using a short-time Fourier transform with a $\unit[10]{ms}$ time window. The two \gls{GW} bursts display distinct spectral evolutions: the first rises from $\sim\unit[550]{Hz}$ to $\sim\unit[1]{kHz}$, while the second steadily hovers near $\unit[1]{kHz}$. The frequencies and time evolution of the most intense ascending spectral feature align with the predicted fundamental quadrupolar $^2f$-mode (green line computed using as independent variable $x=M_{\rm pns}/R_{\rm pns}^3$ instead of the $x=M_{\rm shock}/R_{\rm shock}^3$ in the fits provided by \cite{TorresForne19b}). Moreover, additional  power in the spectrogram growing in parallel to the main mode and resembling higher harmonics of the former one appear  at twice and thrice these frequencies, albeit with progressively lower amplitudes. Both facts strongly suggest that the rising modes in the spectrogram likely result from \gls{PNS} $f$-modes. We stress the fact that the green line in  Figure~\ref{fig:GW_spectro} must be regarded as a proxy of the actual $^2f$-mode frequency. Hence, we refer to this proxy as \emph{approximate} $^2f$ mode.

\begin{figure}
    \centering
    \includegraphics[width=0.5\textwidth]{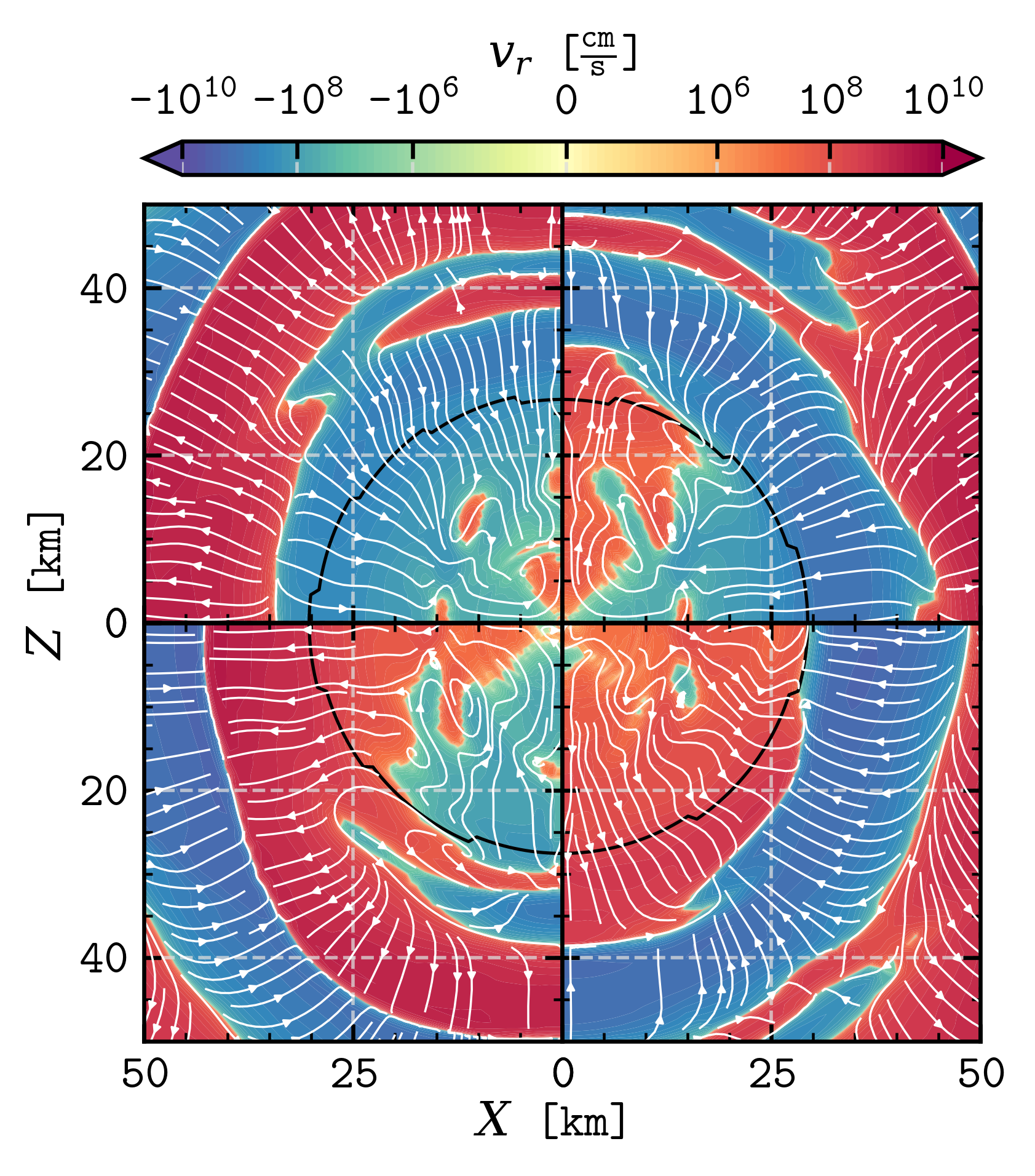}
    \caption{Snapshots of the radial velocity illustrating a full oscillation cycle of the \gls{PNS}  at (clockwise from the top-left panel)  $\unit[1.1283]{s}$,  $\unit[1.1285]{s}$,  $\unit[1.1287]{s}$, and  $\unit[1.1291]{s}$.  Each of the panels shows one hemisphere only.  Black lines represent the \gls{PNS} radius, defined as the isodensity line at $\unit[10^{11}]{g/cm^3}$. Streamlines represent the velocity field. Background colours indicate inward (blue) and outward-moving (red) matter.}
    \label{fig:2D_plot}
\end{figure}

\begin{figure*}
\includegraphics[width=\textwidth]{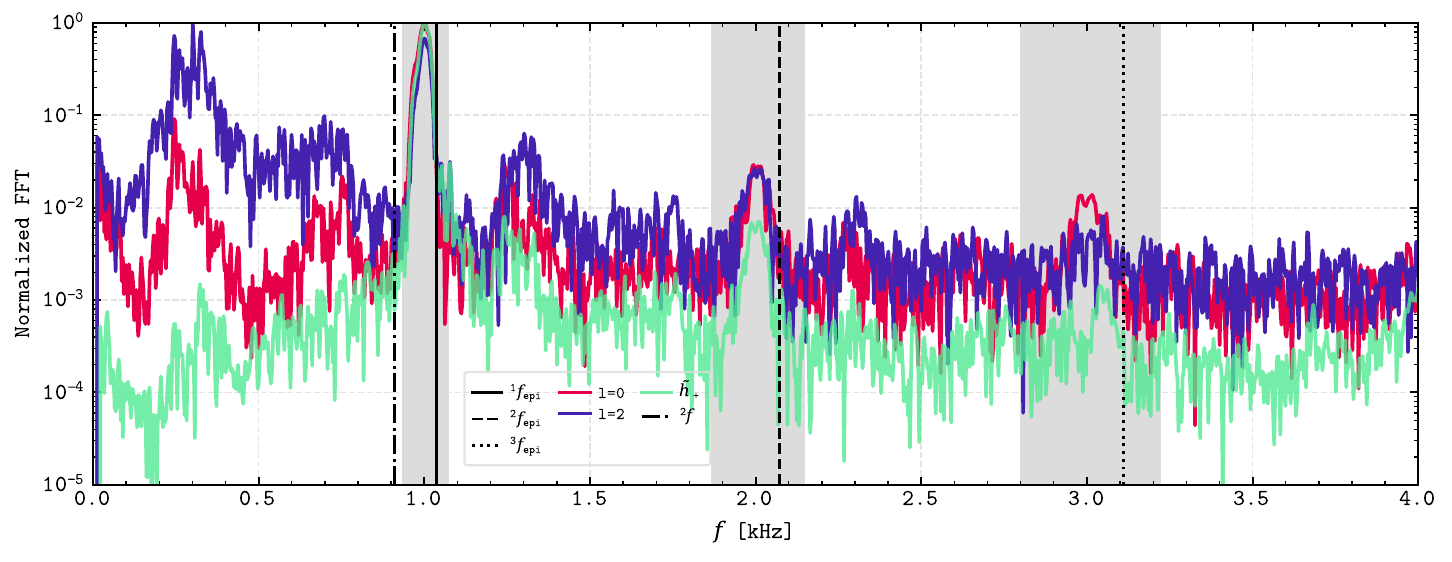} 
\caption{
    \label{fig:epi_fft}
    Normalized Fourier transform of the equatorial density ($l=0$), $v_\theta$ at $\pi/4$ ($l=2$), both outside of taken the \gls{PNS} core outer boundary at $\unit[25]{km}$,
    and of the \gls{GW} signal ($\tilde{h}_+$) for the time interval $\unit[1.1-1.4]{s}$. The solid, dashed, and dotted black vertical lines represent the fundamental frequency and the first and second overtones of the average epicyclic frequency, respectively. Shaded regions
    indicate the uncertainty associated with the frequencies. The dash-dotted line represents the average $^2f$-mode frequency derived with the relations in \cite{TorresForne19}.
}
\end{figure*}

Using the quadrupole formula including surface terms (see, e.g., \cite{Zha_2024PhRvD.110h3034}), we pinpoint the origin of the \gls{GW} strain to the innermost part of the star (hereafter \gls{PNS} core), which we defined as the region within the iso-entropy surface at $\unit[4]{k_B/baryon}$ \citep{Ertl16}, corresponding to a radius of $r_\textnormal{core}\sim\unit[20]{km}$).
The pronounced high-amplitude emission arises from a resonant interplay between \gls{PNS} modes and rotation. 
The centrifugal force causes fluid elements within the \gls{PNS} core to oscillate in the direction of the cylindrical radial distance.
To quantify these oscillations, we use the epicyclic frequency, defined as
\begin{equation}
    \label{eq:epicyclic_frequency_overtone}
    ^nf_\textnormal{epi} = \frac{n}{2\pi}\sqrt{\frac{2\Omega}{R}\frac{\textnormal{d}(\Omega R^2)}{\textnormal{d}R}}, 
\end{equation}
where $n$ is the overtone number, $R$ the cylindrical radius, and $\Omega$ the angular velocity.

Within the \gls{PNS} core, we focus on a conical region spanning the colatitude range $[\unit[15]{^\circ}, \unit[85]{^\circ}]$. At each radius, we determine the maximum epicyclic frequency and subsequently compute the average and standard deviation of these maxima. This specific colatitude range was chosen to exclude the polar region (prone to numerical artifacts in axial symmetry) and the equator, where converging convective flows compromise the accuracy of the $^nf_{\rm epi}$ evaluation. Recognizing that simply taking the maximum may lead to an overestimate, we conservatively define the lower uncertainty as the largest value between the standard deviation and 10\% of the maximum of the epicyclic frequency. The bottom panel of Figure~\ref{fig:GW_spectro} shows the evolution of these averages (blue lines) and uncertainties (shades) for $^1f_\textnormal{epi}$ (solid), $^2f_\textnormal{epi}$ (dashed), and $^3f_\textnormal{epi}$ (dotted).

Resonance occurs when the fundamental epicyclic frequency intersects or comes sufficiently close to the $^2f$-mode frequency. 
To give a first analytic estimate of the latter we employ the quasi-universal relations of \cite{TorresForne19b}. These relations were obtained from non-rotating 1D \gls{CCSN} simulations. Thus, their approximate formulae may  exhibit  deviations from our models, yet they  provide a rough reference for comparison with the epicyclic frequency.

During the resonance phases, the \gls{PNS} core oscillates violently over few tens of milliseconds. Since the \gls{PNS} core and \gls{PNS} essentially coincide during the second resonant phase, the two terms are used interchangeably in the following discussion. In Figure~\ref{fig:2D_plot} we provide a time-sequence visualization of the fluid motion within the \gls{PNS} core and the associated deformation during the second resonant phase, showing the radial velocity ($v_r$) during a full oscillation cycle. The \gls{PNS} core motion takes the form of a large-scale oscillation in which the core undergoes over a full cycle the following dynamical sequence: a roughly isotropic contraction (top left), formation of two vortices, one per hemisphere, with inward/outward radial velocity at the equator/poles (top right), expansion (bottom right), and oppositely oriented vortex pair (bottom left). Over the course of one oscillation, the \gls{PNS} core changes its quadrupole moment due to the differential between motion in the equatorial and polar regions (second and fourth phase). At the outer border of the \gls{PNS} (black line), each cycle of the oscillation launches a wave into the post-shock region, visible in the four panels as the approximately isotropic patterns of alternating positive and negative velocities moving away from the \gls{PNS}.

To further validate the resonance mechanism, we resort to the computed data to give a second estimate of the frequency of the quasi-radial ($l=0$) and quadrupolar ($l=2$) modes. As in \cite{Dimmelmeier2006}, we use as proxies for the amplitudes of the $l=0$ and $l=2$ modes the real parts of the Fourier spectra of the mass density at $r = \unit[25]{km}, \, \theta = \pi/2$ and of $v_{\theta}$ at $r = \unit[25]{km}, \theta = \pi/4$, respectively. The choice of $r=\unit[25]{km}$ ensures that the extracted signal reflects fluid displacements induced by the oscillations of the \gls{PNS} core. In Figure~\ref{fig:epi_fft} we present the \gls{FTA} of the  density and $v_\theta$, each normalized by their maximum during the second resonant phase ($\unit[1.15-1.45]{s}$), along with the \gls{GW} \gls{FTA} computed in the whole domain ($r\le 10^{10}\,$cm).
The \gls{GW} spectrum (green line) exhibits a series of distinct peaks starting at $\sim\unit[1]{kHz}$, spaced at intervals of $\sim\unit[1]{kHz}$. These peaks persist up to $\sim\unit[7]{kHz}$, though their amplitude steadily diminishes with increasing frequency, a behaviour indicative of overtones of the epicyclic frequency.

The spectral peaks of the $l=0$ mode (red line) and $l=2$ (blue line) closely align with those of the \gls{GW} signal and, to a rather good approximation, with the epicyclic frequency and its overtones. 
The slight shift between the local \gls{GW} spectral peak spectrum and the epicyclic frequency is likely due to its overestimation.  
Furthermore, the first peak in the \gls{GW} spectrum, corresponding to the fundamental epicyclic frequency, agrees with the $^2f$-mode estimated via the quasi-universal relations of \cite{TorresForne19}. Together, these findings suggest that the fundamental epicyclic mode is the primary driver of the observed \gls{GW} emission during the resonant phase.

\begin{figure}
\includegraphics[width=0.5\textwidth]{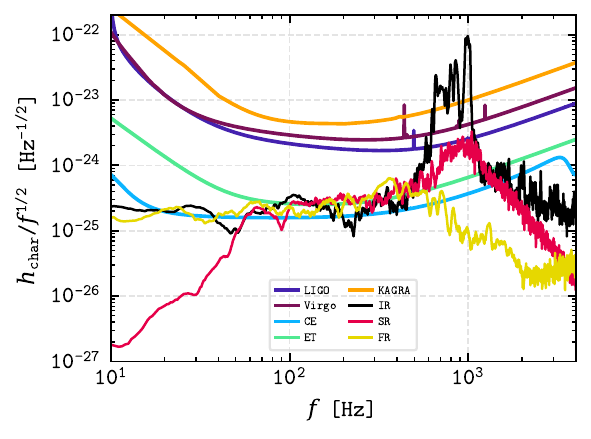}
\caption{
    \label{fig:ASD_comparison}Characteristic \gls{GW} spectra for models \texttt{SR} (red), \texttt{IR} (black), and \texttt{FR} (yellow), assuming a source distance of $\unit[1]{Mpc}$, compared with the design sensitivity curves of current and next-generation interferometers. $h_\texttt{char}$ has been convolved over a window of $\unit[10]{Hz}$ for better visualization.
}
\end{figure}

To assess the detectability of the \gls{GW} signal, we compute its characteristic strain \cite{Moore2015}, $h_\textnormal{char}$,

\begin{equation}
    \label{eq:characteristic_strain}
    h_{\textnormal{char}}(f) = 2 f \, \lvert \tilde{h}(f) \rvert ,
\end{equation}
where $\tilde{h}(f)$ is the Fourier transform of the \gls{GW} strain $h(t)$ after applying a Hann window $w(t)$,
\begin{equation}
    \tilde{h}(f)
    =
    \int_{-\infty}^{+\infty}
    h(t)\, w(t)\, e^{-2\pi i f t}\, \mathrm{d}t .
\end{equation}
The Hann window is defined as
\begin{equation}
    w(t)
    =
    \frac{1}{2}
    \left[
        1 - \cos\!\left( \frac{2\pi t}{T} \right)
    \right],
    \qquad
    0 \le t \le T ,
\end{equation}
and $w(t)=0$ outside the interval $[0,T]$, where $T$ is the total duration
of the signal.

The use of a Hann window reduces spectral leakage associated with the
finite time span of the signal, but it also introduces a reduction of
the Fourier amplitudes. For a coherent signal, the Hann window has a
coherent gain $\langle w \rangle = T^{-1}\!\int_0^T w(t)\,\mathrm{d}t = 1/2$
\cite{Harris_1978}. To preserve the correct normalization of the Fourier
amplitude, and hence of $h_{\textnormal{char}}$, the windowed Fourier
transform is therefore renormalized by the inverse of this coherent
gain. 

In Figure~\ref{fig:ASD_comparison} we show $h_\textnormal{char} / f^{1/2}$ for models \texttt{SR}, \texttt{IR}, and \texttt{FR} at $\mathcal{D} = \unit[1]{Mpc}$, juxtaposed with the \gls{ASD} of current and next-generation interferometers, namely Advanced LIGO \cite{Aasi15,LVK20,LVKASDs}, Advanced Virgo \cite{Acernese15,LVK20,LVKASDs}, KAGRA \cite{Akutsu19,LVK20,LVKASDs}, \gls{ET} \cite{Hild11,ETASD}, and \gls{CE} \cite{Srivastava22,CEASD}.

Model \texttt{SR} notably exhibits a broad low-frequency  hump spanning $\unit[40-90]{Hz}$ alongside a broad high-frequency peak at $\unit[400-1000]{Hz}$. While the low-frequency features fall below the sensitivity range of present and future detectors, the high-frequency peak marginally surpasses the theoretical sensitivity of \gls{CE}. Moreover, in the absence of polar jets,  no significant power is observed below $f \sim\unit[30]{Hz}$.

Model \texttt{FR} exhibits a broad plateau below $\unit[250]{Hz}$.  Its high-frequency peak is shifted down to $\sim\unit[400]{Hz}$ and is markedly weaker compared to models \texttt{SR} and \texttt{IR}. 
Contrary to the previous model, a spectral power excess at low frequencies ($\lesssim\unit[30]{Hz}$), is present, indicating jet formation.

Model \texttt{IR} differs significantly from the others. It displays two low-frequency spectral humps: one at intermediate frequencies $(\unit[60-200]{Hz})$, and another below $\unit[30]{Hz}$.
The most striking feature of \texttt{IR} is the unmistakable resonance imprint on its characteristic strain. This model yields the strongest signal, with a peak strain more than twice that observed in \texttt{SR}, translating into a peak \gls{ASD} more than 10 (100) times larger than in model \texttt{SR} (\texttt{FR})-(see Figure~\ref{fig:ASD_comparison}). 
Four prominent peaks, spanning from $\sim\unit[600]{Hz}$ to $\sim\unit[1]{kHz}$, mark the resonant frequencies during the first and second resonant phases (see Figure~\ref{fig:GW_spectro}). Overtones at $f \gtrsim\unit[2]{kHz}$ and $f \gtrsim\unit[3]{kHz}$ are also present.
To estimate the maximum distance at which the \glspl{GW} from our models remain detectable, we adopt an optimal \gls{SNR} threshold of 12. This value is higher than the usual threshold of 8 used in recent works \cite{Powell2023,Powell2024} to account for the intrinsic overestimation resulting from the axisymmetric nature of the simulation. We define the $\text{SNR}$ as in \citet{Moore2015},
\begin{equation}
    \textnormal{SNR} = \sqrt{\int_{0}^{+\infty}\textnormal{d}\ln f \left(\frac{h_\textnormal{char}(f)}{h_n(f)}\right)^2},
\end{equation}
where $h_n(f) = \sqrt{f}\textnormal{ASD}(f)$, in which we used the \gls{ASD} of the detector noise.
The strong resonant features of model \texttt{IR} lie within the detection range of current interferometers such as Advanced LIGO and Virgo beyond the Andromeda Galaxy \cite{Karachentsev2006}, with a maximum reach of $\unit[1]{Mpc}$. Next-generation observatories like \gls{CE}, with enhanced high-frequency sensitivity, could extend this range to $\sim\unit[7.4]{Mpc}$.

\begin{figure}
\includegraphics[width=0.5\textwidth]{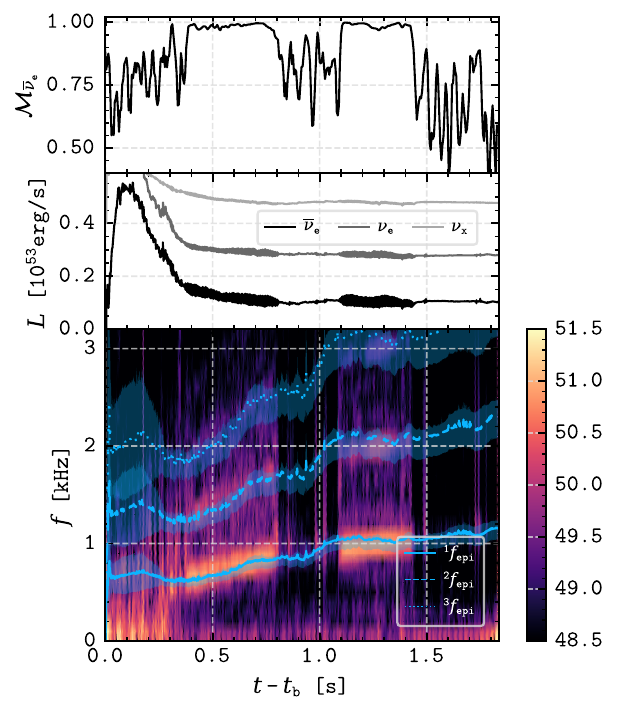} 
\caption{
    \label{fig:Nu_spectro} Top panel: evolution of the matching score between the spectrograms of the  $\overline{\nu}_\textnormal{e}$ and \gls{GW} signals. Middle panel: neutrino luminosities of electron antineutrinos (black line), electron neutrinos (dark gray, shifted by $\unit[0.2\times10^{53}]{erg/s}$), heavy lepton neutrinos (light gray, shifted by $\unit[0.4\times10^{53}]{erg/s}$). Bottom panel: spectrogram of the antineutrino luminosities, with the shaded regions and blue lines denoting the same quantities as in Figure~\ref{fig:GW_spectro}. Both panels refer to model \texttt{IR}.
}
\end{figure}

In model \texttt{IR}, the neutrino emission shows resonant features at frequencies matching those of the \glspl{GW} signal, albeit at slightly lower amplitudes. 
These resonant imprints emerge in the mean energies and luminosities of all three neutrino flavours ($\nu_\textnormal{e}$, 
$\overline{\nu}_\textnormal{e}$, and the combined $\tau-$ and $\mu-$neutrinos, $\nu_\textnormal{x}$).
The two bottom panels of Figure~\ref{fig:Nu_spectro} depict, respectively, the evolution of the global neutrino luminosity (middle panel) and the spectrogram of the electron antineutrino luminosity (bottom panel), both computed with a $\unit[10]{ms}$ time window, analogous to the analysis conducted for the \gls{GW} signal.

During the resonance intervals observed in the \gls{GW} signal ($\unit[0.4-0.8]{s}$ and $\unit[1.1-1.4]{s}$), the global neutrino luminosity oscillates around its mean value. Altough such oscillations are evident in all neutrino flavours, the resonance imprint is most pronounced in the $\overline{\nu}_\textnormal{e}$ component, followed by $\nu_\textnormal{e}$, and is least apparent in $\nu_\textnormal{x}$ signal.
The spectrogram of the $\overline{\nu}_\textnormal{e}$ luminosity (bottom panel of Figure~\ref{fig:Nu_spectro}) mirrors that of the \glspl{GW}, displaying two distinct phases of robust, resonantly excited oscillations at the epicyclic frequency or its overtones (blue lines).

To quantify the correlation between the \gls{GW} and neutrino signals, we use the matching score \cite{Suvorova19},
\begin{equation}
    \label{eq:matching}
    \mathcal{M}_{\nu_i}(t) = \frac{\langle S_\textnormal{GW}(f, t) | S_{\nu_i}(f,t) \rangle}{\sqrt{\langle S_\textnormal{GW}(f, t) | S_\textnormal{GW}(f, t) \rangle \langle S_{\nu_i}(f,t) | S_{\nu_i}(f,t) \rangle}},
\end{equation}
where $S_j(f, t)$ ($j=\textnormal{GW},\, \nu_i$) denotes the magnitude of the spectrogram computed with a $\unit[10]{ms}$ time window, and $\langle\cdot|\cdot\rangle$ represents the inner product in frequency space.

The top panel of Figure~\ref{fig:Nu_spectro} depicts the evolution of the matching score between the \glspl{GW} (bottom panel of Figure~\ref{fig:GW_spectro}) and $\overline{\nu}_\textnormal{e}$ spectrograms (bottom panel of Figure~\ref{fig:Nu_spectro}; results for the other flavours are similar) for frequencies above $\unit[200]{Hz}$, a threshold set to minimize stochastic contributions from convection and \gls{SASI}. Throughout the simulation, the matching score consistently exceeds 0.6. Notably, during the two resonant phases ($\unit[0.4-0.8]{s}$ and $\unit[1.1-1.4]{s}$), it approaches unity, unequivocally demonstrating that the amplitude modulations  in both signals are correlated and originate from the same rotationally driven oscillations of the \gls{PNS} core.

\section{Discussion and Conclusions}
\label{sec:conclusion}
We have reported results from \gls{2D} \gls{CCSN} simulations for a model with a pre-bounce central rotation rate of $\Omega_{\rm c} \sim \unit[1]{rad/s}$ that reveal extended intervals—lasting several hundred milliseconds—of extraordinarily strong \gls{GW} emission. Such pronounced signals are absent in otherwise identical progenitors with either slower or faster initial rotation. We attribute this enhanced emission to a resonance between the frequency of the fundamental quadrupolar $^2f$-mode and the frequency of epicyclic oscillations at the outer boundary of the \gls{PNS} inner core.
Notably, while the \gls{GW} amplitudes are comparable to the bounce signals observed in rapidly rotating cores, they persist for much longer. Crucially, the resonance signatures dominate the most prominent features of the characteristic strain, and their typical frequencies of around $\unit[1]{kHz}$ renders them compelling targets for ground-based \gls{GW} detectors. Although higher harmonics are also excited, their detection is impeded by both their high frequencies and lower amplitudes. We also detect imprints of these resonant oscillations in the neutrino emission from the core, with luminosities and mean energies modulated at the same frequencies, most prominently in the $\bar{\nu}_{e}$ and $\nu_e$ channels. 

While our findings suggest a new effect relevant for multimessenger observations of \glspl{CCSN}, further investigation is clearly warranted. In realistic \gls{CCSN} cores, deviations from axisymmetry arise rapidly; at the rotation rates explored here, non-axisymmetric instabilities may develop and alter the rotational profile, potentially modifying the conditions for resonance. These effects must be rigorously examined using full \gls{3D} simulations. Moreover, \gls{2D} simulations may
overestimate \gls{GW} amplitudes compared to \gls{3D} \cite{Murphy24}, which represents a limitation of the present study. Furthermore, future studies should probe whether this resonance is sensitive to other progenitor properties,  such as initial mass or metallicity, and if it can offer new constraints on the nuclear \gls{EOS}.

Bearing in mind these caveats, the pronounced, long-lasting modulation of the \gls{GW} signal induced by resonance reported in this work could represent a breakthrough for multimessenger detections of \glspl{CCSN}.
Given the substantial \gls{GW} amplitude enhancement, this resonant signature should remain detectable out to distances of $\sim\unit[1]{Mpc}$ with current detectors. With \gls{CE}, the detection horizon extends to $\sim\unit[7.4]{Mpc}$. This makes the signal a compelling target for both current and next-generation interferometers.

To provide an estimate of the expected number of detectable events, we
approximate the detection rate as
\begin{equation}
\label{eq:rate_estimate}
\mathcal{R}_{\rm det}
\;\simeq\;
\mathcal{R}_{\rm CCSN}(D<\unit[10]{Mpc})\,V(d_{\rm r})\,f_{\rm rot},
\end{equation}
where $\mathcal{R}_{\rm CCSN}(D<\unit[10]{Mpc})$ is the local volumetric core-collapse supernova rate (i.e., within a distance $D<\unit[10]{Mpc}$),
$V(d_{\rm r})$ is the accessible volume set by the detector horizon distance
$d_{\rm r}$, and $f_{\rm rot}$ is the fraction of massive-star progenitors that meet
the (model-dependent) conditions for the resonance (e.g. sufficiently rapid
presupernova core rotation; namely $\sim \unit[1]{rad/s}$ \cite{Holgado22}; see Appendix~\ref{app:rotation}). We take $V(d_{\rm r})=4\pi d_{\rm r}^3/3$,
and adopt $d_{\rm r}\simeq \unit[7.4]{Mpc}$ for CE (as inferred from our SNR analysis).
For the local CCSN rate we use the commonly employed empirical estimate of
$\sim 1 - 2$ CCSN per year within $\unit[10]{Mpc}$, with uncertainties dominated by incompleteness due to dust extinction and intrinsically faint events (e.g., \cite{Ando_2004ApJ...607...20, Botticella_2012A&A...537A.132, Horiuchi_2013ApJ...769..113, Capellaro_2015IAUS..317..181}). That corresponds to
$\mathcal{R}_{\rm CCSN} (D<\unit[10]{Mpc})\;\approx\;
2.4\times 10^{-4}\ {\rm yr}^{-1}\,{\rm Mpc}^{-3}$.
Stellar-evolution and population-synthesis studies indicate that only a small fraction of massive-star progenitors are expected to retain presupernova core rotation rates of order $
\sim \unit[1]{rad/s}$, with estimates ranging from $f_{\rm rot}\sim 10^{-2}$ up to 
$\sim 10^{-1}$ in optimistic scenarios involving chemically homogeneous evolution or strong binary interaction (e.g., \cite{Heger_2005ApJ...626..350, AguileraDena18, 
Holgado22}). As a fiducial, but highly uncertain value one may set $f_{\rm rot}\!=\!0.03$, yielding
\begin{equation}
\begin{aligned}
\mathcal{R}_{\rm det}\;\simeq\;
0.012~{\rm yr}^{-1}
&\left(\frac{f_{\rm rot}}{0.03}\right)\times
\left(\frac{d_{\rm hor}}{7.4~{\rm Mpc}}\right)^3
\times \\
&\left(\frac{\mathcal{R}_{\rm CCSN}(D<\unit[10]{Mpc})}{2.4\times10^{-4}~{\rm yr}^{-1}\,{\rm Mpc}^{-3}}\right),
\end{aligned}
\end{equation}
i.e., of order one event every $\sim 80$\,years.

\begin{acknowledgments}
      \textit{Acknowledgements.} We acknowledge support through the grants PID2021-127495NB-I00 and PID2024-159689NB-C21 funded by MICIU/AEI/10.13039/501100011033 and by FEDER/EU, and the Astrophysics and High Energy Physics programme of the Generalitat Valenciana ASFAE/2022/026 funded by MCIN and the European Union NextGenerationEU (PRTR-C17.I1) as well as support from the Prometeo excellence programme grants CIPROM/2022/13 and CIPROM/2022/49 funded by the Generalitat Valenciana.
      MO was supported by the Ramón y Cajal programme of the Agencia Estatal de Investigación (RYC2018-024938-I).
      MC acknowledges the support through the Generalitat Valenciana via the grant CIDEGENT/2019/031.
      The computations have been performed on servers Lluisvives and Tirant-4 (grant AECT-2025-2-0002) of the Servei d'Informàtica de la Universitat de València and on the Red Española de Supercomputación (RES) on MareNostrum (grants AECT-2025-1-0012 and AECT-2025-2-0006).
\end{acknowledgments}

\bibliography{bibliography}

\begin{thebibliography}{65}%
\makeatletter
\providecommand \@ifxundefined [1]{%
 \@ifx{#1\undefined}
}%
\providecommand \@ifnum [1]{%
 \ifnum #1\expandafter \@firstoftwo
 \else \expandafter \@secondoftwo
 \fi
}%
\providecommand \@ifx [1]{%
 \ifx #1\expandafter \@firstoftwo
 \else \expandafter \@secondoftwo
 \fi
}%
\providecommand \natexlab [1]{#1}%
\providecommand \enquote  [1]{``#1''}%
\providecommand \bibnamefont  [1]{#1}%
\providecommand \bibfnamefont [1]{#1}%
\providecommand \citenamefont [1]{#1}%
\providecommand \href@noop [0]{\@secondoftwo}%
\providecommand \href [0]{\begingroup \@sanitize@url \@href}%
\providecommand \@href[1]{\@@startlink{#1}\@@href}%
\providecommand \@@href[1]{\endgroup#1\@@endlink}%
\providecommand \@sanitize@url [0]{\catcode `\\12\catcode `\$12\catcode `\&12\catcode `\#12\catcode `\^12\catcode `\_12\catcode `\%12\relax}%
\providecommand \@@startlink[1]{}%
\providecommand \@@endlink[0]{}%
\providecommand \url  [0]{\begingroup\@sanitize@url \@url }%
\providecommand \@url [1]{\endgroup\@href {#1}{\urlprefix }}%
\providecommand \urlprefix  [0]{URL }%
\providecommand \Eprint [0]{\href }%
\providecommand \doibase [0]{https://doi.org/}%
\providecommand \selectlanguage [0]{\@gobble}%
\providecommand \bibinfo  [0]{\@secondoftwo}%
\providecommand \bibfield  [0]{\@secondoftwo}%
\providecommand \translation [1]{[#1]}%
\providecommand \BibitemOpen [0]{}%
\providecommand \bibitemStop [0]{}%
\providecommand \bibitemNoStop [0]{.\EOS\space}%
\providecommand \EOS [0]{\spacefactor3000\relax}%
\providecommand \BibitemShut  [1]{\csname bibitem#1\endcsname}%
\let\auto@bib@innerbib\@empty
\bibitem [{\citenamefont {{Bionta}}\ \emph {et~al.}(1987)\citenamefont {{Bionta}}, \citenamefont {{Blewitt}}, \citenamefont {{Bratton}}, \citenamefont {{Casper}}, \citenamefont {{Ciocio}}, \citenamefont {{Claus}}, \citenamefont {{Cortez}}, \citenamefont {{Crouch}}, \citenamefont {{Dye}}, \citenamefont {{Errede}}, \citenamefont {{Foster}}, \citenamefont {{Gajewski}}, \citenamefont {{Ganezer}}, \citenamefont {{Goldhaber}}, \citenamefont {{Haines}}, \citenamefont {{Jones}}, \citenamefont {{Kielczewska}}, \citenamefont {{Kropp}}, \citenamefont {{Learned}}, \citenamefont {{Losecco}}, \citenamefont {{Matthews}}, \citenamefont {{Miller}}, \citenamefont {{Mudan}}, \citenamefont {{Park}}, \citenamefont {{Price}}, \citenamefont {{Reines}}, \citenamefont {{Schultz}}, \citenamefont {{Seidel}}, \citenamefont {{Shumard}}, \citenamefont {{Sinclair}}, \citenamefont {{Sobel}}, \citenamefont {{Stone}}, \citenamefont {{Sulak}}, \citenamefont {{Svoboda}}, \citenamefont {{Thornton}}, \citenamefont {{van der Velde}},\ and\
  \citenamefont {{Wuest}}}]{Bionta87}%
  \BibitemOpen
  \bibfield  {author} {\bibinfo {author} {\bibfnamefont {R.~M.}\ \bibnamefont {{Bionta}}}, \bibinfo {author} {\bibfnamefont {G.}~\bibnamefont {{Blewitt}}}, \bibinfo {author} {\bibfnamefont {C.~B.}\ \bibnamefont {{Bratton}}}, \bibinfo {author} {\bibfnamefont {D.}~\bibnamefont {{Casper}}}, \bibinfo {author} {\bibfnamefont {A.}~\bibnamefont {{Ciocio}}}, \bibinfo {author} {\bibfnamefont {R.}~\bibnamefont {{Claus}}}, \bibinfo {author} {\bibfnamefont {B.}~\bibnamefont {{Cortez}}}, \bibinfo {author} {\bibfnamefont {M.}~\bibnamefont {{Crouch}}}, \bibinfo {author} {\bibfnamefont {S.~T.}\ \bibnamefont {{Dye}}}, \bibinfo {author} {\bibfnamefont {S.}~\bibnamefont {{Errede}}}, \bibinfo {author} {\bibfnamefont {G.~W.}\ \bibnamefont {{Foster}}}, \bibinfo {author} {\bibfnamefont {W.}~\bibnamefont {{Gajewski}}}, \bibinfo {author} {\bibfnamefont {K.~S.}\ \bibnamefont {{Ganezer}}}, \bibinfo {author} {\bibfnamefont {M.}~\bibnamefont {{Goldhaber}}}, \bibinfo {author} {\bibfnamefont {T.~J.}\ \bibnamefont {{Haines}}}, \bibinfo
  {author} {\bibfnamefont {T.~W.}\ \bibnamefont {{Jones}}}, \bibinfo {author} {\bibfnamefont {D.}~\bibnamefont {{Kielczewska}}}, \bibinfo {author} {\bibfnamefont {W.~R.}\ \bibnamefont {{Kropp}}}, \bibinfo {author} {\bibfnamefont {J.~G.}\ \bibnamefont {{Learned}}}, \bibinfo {author} {\bibfnamefont {J.~M.}\ \bibnamefont {{Losecco}}}, \bibinfo {author} {\bibfnamefont {J.}~\bibnamefont {{Matthews}}}, \bibinfo {author} {\bibfnamefont {R.}~\bibnamefont {{Miller}}}, \bibinfo {author} {\bibfnamefont {M.~S.}\ \bibnamefont {{Mudan}}}, \bibinfo {author} {\bibfnamefont {H.~S.}\ \bibnamefont {{Park}}}, \bibinfo {author} {\bibfnamefont {L.~R.}\ \bibnamefont {{Price}}}, \bibinfo {author} {\bibfnamefont {F.}~\bibnamefont {{Reines}}}, \bibinfo {author} {\bibfnamefont {J.}~\bibnamefont {{Schultz}}}, \bibinfo {author} {\bibfnamefont {S.}~\bibnamefont {{Seidel}}}, \bibinfo {author} {\bibfnamefont {E.}~\bibnamefont {{Shumard}}}, \bibinfo {author} {\bibfnamefont {D.}~\bibnamefont {{Sinclair}}}, \bibinfo {author} {\bibfnamefont
  {H.~W.}\ \bibnamefont {{Sobel}}}, \bibinfo {author} {\bibfnamefont {J.~L.}\ \bibnamefont {{Stone}}}, \bibinfo {author} {\bibfnamefont {L.~R.}\ \bibnamefont {{Sulak}}}, \bibinfo {author} {\bibfnamefont {R.}~\bibnamefont {{Svoboda}}}, \bibinfo {author} {\bibfnamefont {G.}~\bibnamefont {{Thornton}}}, \bibinfo {author} {\bibfnamefont {J.~C.}\ \bibnamefont {{van der Velde}}},\ and\ \bibinfo {author} {\bibfnamefont {C.}~\bibnamefont {{Wuest}}},\ }\bibfield  {title} {\bibinfo {title} {{Observation of a neutrino burst in coincidence with supernova 1987A in the Large Magellanic Cloud}},\ }\href {https://doi.org/10.1103/PhysRevLett.58.1494} {\bibfield  {journal} {\bibinfo  {journal} {\prl}\ }\textbf {\bibinfo {volume} {58}},\ \bibinfo {pages} {1494} (\bibinfo {year} {1987})}\BibitemShut {NoStop}%
\bibitem [{\citenamefont {{Hirata}}\ \emph {et~al.}(1987)\citenamefont {{Hirata}}, \citenamefont {{Kajita}}, \citenamefont {{Koshiba}}, \citenamefont {{Nakahata}}, \citenamefont {{Oyama}}, \citenamefont {{Sato}}, \citenamefont {{Suzuki}}, \citenamefont {{Takita}}, \citenamefont {{Totsuka}}, \citenamefont {{Kifune}}, \citenamefont {{Suda}}, \citenamefont {{Takahashi}}, \citenamefont {{Tanimori}}, \citenamefont {{Miyano}}, \citenamefont {{Yamada}}, \citenamefont {{Beier}}, \citenamefont {{Feldscher}}, \citenamefont {{Kim}}, \citenamefont {{Mann}}, \citenamefont {{Newcomer}}, \citenamefont {{van}}, \citenamefont {{Zhang}},\ and\ \citenamefont {{Cortez}}}]{Hirata87}%
  \BibitemOpen
  \bibfield  {author} {\bibinfo {author} {\bibfnamefont {K.}~\bibnamefont {{Hirata}}}, \bibinfo {author} {\bibfnamefont {T.}~\bibnamefont {{Kajita}}}, \bibinfo {author} {\bibfnamefont {M.}~\bibnamefont {{Koshiba}}}, \bibinfo {author} {\bibfnamefont {M.}~\bibnamefont {{Nakahata}}}, \bibinfo {author} {\bibfnamefont {Y.}~\bibnamefont {{Oyama}}}, \bibinfo {author} {\bibfnamefont {N.}~\bibnamefont {{Sato}}}, \bibinfo {author} {\bibfnamefont {A.}~\bibnamefont {{Suzuki}}}, \bibinfo {author} {\bibfnamefont {M.}~\bibnamefont {{Takita}}}, \bibinfo {author} {\bibfnamefont {Y.}~\bibnamefont {{Totsuka}}}, \bibinfo {author} {\bibfnamefont {T.}~\bibnamefont {{Kifune}}}, \bibinfo {author} {\bibfnamefont {T.}~\bibnamefont {{Suda}}}, \bibinfo {author} {\bibfnamefont {K.}~\bibnamefont {{Takahashi}}}, \bibinfo {author} {\bibfnamefont {T.}~\bibnamefont {{Tanimori}}}, \bibinfo {author} {\bibfnamefont {K.}~\bibnamefont {{Miyano}}}, \bibinfo {author} {\bibfnamefont {M.}~\bibnamefont {{Yamada}}}, \bibinfo {author} {\bibfnamefont
  {E.~W.}\ \bibnamefont {{Beier}}}, \bibinfo {author} {\bibfnamefont {L.~R.}\ \bibnamefont {{Feldscher}}}, \bibinfo {author} {\bibfnamefont {S.~B.}\ \bibnamefont {{Kim}}}, \bibinfo {author} {\bibfnamefont {A.~K.}\ \bibnamefont {{Mann}}}, \bibinfo {author} {\bibfnamefont {F.~M.}\ \bibnamefont {{Newcomer}}}, \bibinfo {author} {\bibfnamefont {R.}~\bibnamefont {{van}}}, \bibinfo {author} {\bibfnamefont {W.}~\bibnamefont {{Zhang}}},\ and\ \bibinfo {author} {\bibfnamefont {B.~G.}\ \bibnamefont {{Cortez}}},\ }\bibfield  {title} {\bibinfo {title} {{Observation of a neutrino burst from the supernova SN1987A}},\ }\href {https://doi.org/10.1103/PhysRevLett.58.1490} {\bibfield  {journal} {\bibinfo  {journal} {\prl}\ }\textbf {\bibinfo {volume} {58}},\ \bibinfo {pages} {1490} (\bibinfo {year} {1987})}\BibitemShut {NoStop}%
\bibitem [{\citenamefont {{Alexeyev}}\ \emph {et~al.}(1988)\citenamefont {{Alexeyev}}, \citenamefont {{Alexeyeva}}, \citenamefont {{Krivosheina}},\ and\ \citenamefont {{Volchenko}}}]{Alexeyev88}%
  \BibitemOpen
  \bibfield  {author} {\bibinfo {author} {\bibfnamefont {E.~N.}\ \bibnamefont {{Alexeyev}}}, \bibinfo {author} {\bibfnamefont {L.~N.}\ \bibnamefont {{Alexeyeva}}}, \bibinfo {author} {\bibfnamefont {I.~V.}\ \bibnamefont {{Krivosheina}}},\ and\ \bibinfo {author} {\bibfnamefont {V.~I.}\ \bibnamefont {{Volchenko}}},\ }\bibfield  {title} {\bibinfo {title} {{Detection of the neutrino signal from SN 1987A in the LMC using the INR Baksan underground scintillation telescope}},\ }\href {https://doi.org/10.1016/0370-2693(88)91651-6} {\bibfield  {journal} {\bibinfo  {journal} {Physics Letters B}\ }\textbf {\bibinfo {volume} {205}},\ \bibinfo {pages} {209} (\bibinfo {year} {1988})}\BibitemShut {NoStop}%
\bibitem [{\citenamefont {{Dotani}}\ \emph {et~al.}(1987)\citenamefont {{Dotani}}, \citenamefont {{Hayashida}}, \citenamefont {{Inoue}}, \citenamefont {{Itoh}}, \citenamefont {{Koyama}}, \citenamefont {{Makino}}, \citenamefont {{Mitsuda}}, \citenamefont {{Murakami}}, \citenamefont {{Oda}}, \citenamefont {{Ogawara}}, \citenamefont {{Takano}}, \citenamefont {{Tanaka}}, \citenamefont {{Yoshida}}, \citenamefont {{Makishima}}, \citenamefont {{Ohashi}}, \citenamefont {{Kawai}}, \citenamefont {{Matsuoka}}, \citenamefont {{Hoshi}}, \citenamefont {{Hayakawa}}, \citenamefont {{Kii}}, \citenamefont {{Kunieda}}, \citenamefont {{Nagase}}, \citenamefont {{Tawara}}, \citenamefont {{Hatsukade}}, \citenamefont {{Kitamoto}}, \citenamefont {{Miyamoto}}, \citenamefont {{Tsunemi}}, \citenamefont {{Yamashita}}, \citenamefont {{Nakagawa}}, \citenamefont {{Yamauchi}}, \citenamefont {{Turner}}, \citenamefont {{Pounds}}, \citenamefont {{Thomas}}, \citenamefont {{Stewart}}, \citenamefont {{Cruise}}, \citenamefont {{Patchett}},\ and\
  \citenamefont {{Reading}}}]{Dotani87}%
  \BibitemOpen
  \bibfield  {author} {\bibinfo {author} {\bibfnamefont {T.}~\bibnamefont {{Dotani}}}, \bibinfo {author} {\bibfnamefont {K.}~\bibnamefont {{Hayashida}}}, \bibinfo {author} {\bibfnamefont {H.}~\bibnamefont {{Inoue}}}, \bibinfo {author} {\bibfnamefont {M.}~\bibnamefont {{Itoh}}}, \bibinfo {author} {\bibfnamefont {K.}~\bibnamefont {{Koyama}}}, \bibinfo {author} {\bibfnamefont {F.}~\bibnamefont {{Makino}}}, \bibinfo {author} {\bibfnamefont {K.}~\bibnamefont {{Mitsuda}}}, \bibinfo {author} {\bibfnamefont {T.}~\bibnamefont {{Murakami}}}, \bibinfo {author} {\bibfnamefont {M.}~\bibnamefont {{Oda}}}, \bibinfo {author} {\bibfnamefont {Y.}~\bibnamefont {{Ogawara}}}, \bibinfo {author} {\bibfnamefont {S.}~\bibnamefont {{Takano}}}, \bibinfo {author} {\bibfnamefont {Y.}~\bibnamefont {{Tanaka}}}, \bibinfo {author} {\bibfnamefont {A.}~\bibnamefont {{Yoshida}}}, \bibinfo {author} {\bibfnamefont {K.}~\bibnamefont {{Makishima}}}, \bibinfo {author} {\bibfnamefont {T.}~\bibnamefont {{Ohashi}}}, \bibinfo {author} {\bibfnamefont
  {N.}~\bibnamefont {{Kawai}}}, \bibinfo {author} {\bibfnamefont {M.}~\bibnamefont {{Matsuoka}}}, \bibinfo {author} {\bibfnamefont {R.}~\bibnamefont {{Hoshi}}}, \bibinfo {author} {\bibfnamefont {S.}~\bibnamefont {{Hayakawa}}}, \bibinfo {author} {\bibfnamefont {T.}~\bibnamefont {{Kii}}}, \bibinfo {author} {\bibfnamefont {H.}~\bibnamefont {{Kunieda}}}, \bibinfo {author} {\bibfnamefont {F.}~\bibnamefont {{Nagase}}}, \bibinfo {author} {\bibfnamefont {Y.}~\bibnamefont {{Tawara}}}, \bibinfo {author} {\bibfnamefont {I.}~\bibnamefont {{Hatsukade}}}, \bibinfo {author} {\bibfnamefont {S.}~\bibnamefont {{Kitamoto}}}, \bibinfo {author} {\bibfnamefont {S.}~\bibnamefont {{Miyamoto}}}, \bibinfo {author} {\bibfnamefont {H.}~\bibnamefont {{Tsunemi}}}, \bibinfo {author} {\bibfnamefont {K.}~\bibnamefont {{Yamashita}}}, \bibinfo {author} {\bibfnamefont {M.}~\bibnamefont {{Nakagawa}}}, \bibinfo {author} {\bibfnamefont {M.}~\bibnamefont {{Yamauchi}}}, \bibinfo {author} {\bibfnamefont {M.~J.~L.}\ \bibnamefont {{Turner}}}, \bibinfo
  {author} {\bibfnamefont {K.~A.}\ \bibnamefont {{Pounds}}}, \bibinfo {author} {\bibfnamefont {H.~D.}\ \bibnamefont {{Thomas}}}, \bibinfo {author} {\bibfnamefont {G.~C.}\ \bibnamefont {{Stewart}}}, \bibinfo {author} {\bibfnamefont {A.~M.}\ \bibnamefont {{Cruise}}}, \bibinfo {author} {\bibfnamefont {B.~E.}\ \bibnamefont {{Patchett}}},\ and\ \bibinfo {author} {\bibfnamefont {D.~H.}\ \bibnamefont {{Reading}}},\ }\bibfield  {title} {\bibinfo {title} {{Discovery of an unusual hard X-ray source in the region of supernova 1987A}},\ }\href {https://doi.org/10.1038/330230a0} {\bibfield  {journal} {\bibinfo  {journal} {\nat}\ }\textbf {\bibinfo {volume} {330}},\ \bibinfo {pages} {230} (\bibinfo {year} {1987})}\BibitemShut {NoStop}%
\bibitem [{\citenamefont {{Matz}}\ \emph {et~al.}(1988)\citenamefont {{Matz}}, \citenamefont {{Share}}, \citenamefont {{Leising}}, \citenamefont {{Chupp}}, \citenamefont {{Vestrand}}, \citenamefont {{Purcell}}, \citenamefont {{Strickman}},\ and\ \citenamefont {{Reppin}}}]{Matz88}%
  \BibitemOpen
  \bibfield  {author} {\bibinfo {author} {\bibfnamefont {S.~M.}\ \bibnamefont {{Matz}}}, \bibinfo {author} {\bibfnamefont {G.~H.}\ \bibnamefont {{Share}}}, \bibinfo {author} {\bibfnamefont {M.~D.}\ \bibnamefont {{Leising}}}, \bibinfo {author} {\bibfnamefont {E.~L.}\ \bibnamefont {{Chupp}}}, \bibinfo {author} {\bibfnamefont {W.~T.}\ \bibnamefont {{Vestrand}}}, \bibinfo {author} {\bibfnamefont {W.~R.}\ \bibnamefont {{Purcell}}}, \bibinfo {author} {\bibfnamefont {M.~S.}\ \bibnamefont {{Strickman}}},\ and\ \bibinfo {author} {\bibfnamefont {C.}~\bibnamefont {{Reppin}}},\ }\bibfield  {title} {\bibinfo {title} {{Gamma-ray line emission from SN1987A}},\ }\href {https://doi.org/10.1038/331416a0} {\bibfield  {journal} {\bibinfo  {journal} {\nat}\ }\textbf {\bibinfo {volume} {331}},\ \bibinfo {pages} {416} (\bibinfo {year} {1988})}\BibitemShut {NoStop}%
\bibitem [{\citenamefont {{Scholberg}}(2012)}]{Scholberg12}%
  \BibitemOpen
  \bibfield  {author} {\bibinfo {author} {\bibfnamefont {K.}~\bibnamefont {{Scholberg}}},\ }\bibfield  {title} {\bibinfo {title} {{Supernova Neutrino Detection}},\ }\href {https://doi.org/10.1146/annurev-nucl-102711-095006} {\bibfield  {journal} {\bibinfo  {journal} {Annual Review of Nuclear and Particle Science}\ }\textbf {\bibinfo {volume} {62}},\ \bibinfo {pages} {81} (\bibinfo {year} {2012})}\BibitemShut {NoStop}%
\bibitem [{\citenamefont {{The IceCube Collaboration}}(2017)}]{IceCube}%
  \BibitemOpen
  \bibfield  {author} {\bibinfo {author} {\bibnamefont {{The IceCube Collaboration}}},\ }\bibfield  {title} {\bibinfo {title} {{The IceCube Neutrino Observatory: instrumentation and online systems}},\ }\href {https://doi.org/10.1088/1748-0221/12/03/P03012} {\bibfield  {journal} {\bibinfo  {journal} {Journal of Instrumentation}\ }\textbf {\bibinfo {volume} {12}}\bibinfo  {number} { (3)},\ \bibinfo {pages} {P03012}}\BibitemShut {NoStop}%
\bibitem [{\citenamefont {{KM3NeT Collaboration}}(2016)}]{KM3NeT}%
  \BibitemOpen
\bibfield  {number} {  }\bibfield  {author} {\bibinfo {author} {\bibnamefont {{KM3NeT Collaboration}}},\ }\bibfield  {title} {\bibinfo {title} {{Letter of intent for KM3NeT 2.0}},\ }\href {https://doi.org/10.1088/0954-3899/43/8/084001} {\bibfield  {journal} {\bibinfo  {journal} {Journal of Physics G Nuclear Physics}\ }\textbf {\bibinfo {volume} {43}},\ \bibinfo {eid} {084001} (\bibinfo {year} {2016})}\BibitemShut {NoStop}%
\bibitem [{\citenamefont {{Suwa}}\ \emph {et~al.}(2022)\citenamefont {{Suwa}}, \citenamefont {{Harada}}, \citenamefont {{Harada}}, \citenamefont {{Koshio}}, \citenamefont {{Mori}}, \citenamefont {{Nakanishi}}, \citenamefont {{Nakazato}}, \citenamefont {{Sumiyoshi}},\ and\ \citenamefont {{Wendell}}}]{Suwa22}%
  \BibitemOpen
  \bibfield  {author} {\bibinfo {author} {\bibfnamefont {Y.}~\bibnamefont {{Suwa}}}, \bibinfo {author} {\bibfnamefont {A.}~\bibnamefont {{Harada}}}, \bibinfo {author} {\bibfnamefont {M.}~\bibnamefont {{Harada}}}, \bibinfo {author} {\bibfnamefont {Y.}~\bibnamefont {{Koshio}}}, \bibinfo {author} {\bibfnamefont {M.}~\bibnamefont {{Mori}}}, \bibinfo {author} {\bibfnamefont {F.}~\bibnamefont {{Nakanishi}}}, \bibinfo {author} {\bibfnamefont {K.}~\bibnamefont {{Nakazato}}}, \bibinfo {author} {\bibfnamefont {K.}~\bibnamefont {{Sumiyoshi}}},\ and\ \bibinfo {author} {\bibfnamefont {R.~A.}\ \bibnamefont {{Wendell}}},\ }\bibfield  {title} {\bibinfo {title} {{Observing Supernova Neutrino Light Curves with Super-Kamiokande. III. Extraction of Mass and Radius of Neutron Stars from Synthetic Data}},\ }\href {https://doi.org/10.3847/1538-4357/ac795e} {\bibfield  {journal} {\bibinfo  {journal} {\apj}\ }\textbf {\bibinfo {volume} {934}},\ \bibinfo {eid} {15} (\bibinfo {year} {2022})}\BibitemShut {NoStop}%
\bibitem [{\citenamefont {{Mart{\'\i}nez-Mirav{\'e}}}\ \emph {et~al.}(2024)\citenamefont {{Mart{\'\i}nez-Mirav{\'e}}}, \citenamefont {{Tamborra}}, \citenamefont {{Aloy}},\ and\ \citenamefont {{Obergaulinger}}}]{Martinez-Mirave_2024PhRvD.110j3029}%
  \BibitemOpen
  \bibfield  {author} {\bibinfo {author} {\bibfnamefont {P.}~\bibnamefont {{Mart{\'\i}nez-Mirav{\'e}}}}, \bibinfo {author} {\bibfnamefont {I.}~\bibnamefont {{Tamborra}}}, \bibinfo {author} {\bibfnamefont {M.~{\'A}.}\ \bibnamefont {{Aloy}}},\ and\ \bibinfo {author} {\bibfnamefont {M.}~\bibnamefont {{Obergaulinger}}},\ }\bibfield  {title} {\bibinfo {title} {{Diffuse neutrino background from magnetorotational stellar core collapses}},\ }\href {https://doi.org/10.1103/PhysRevD.110.103029} {\bibfield  {journal} {\bibinfo  {journal} {\prd}\ }\textbf {\bibinfo {volume} {110}},\ \bibinfo {eid} {103029} (\bibinfo {year} {2024})}\BibitemShut {NoStop}%
\bibitem [{\citenamefont {{LIGO Scientific Collaboration}}(2015)}]{Aasi15}%
  \BibitemOpen
  \bibfield  {author} {\bibinfo {author} {\bibnamefont {{LIGO Scientific Collaboration}}},\ }\bibfield  {title} {\bibinfo {title} {{Advanced LIGO}},\ }\href {https://doi.org/10.1088/0264-9381/32/7/074001} {\bibfield  {journal} {\bibinfo  {journal} {Classical and Quantum Gravity}\ }\textbf {\bibinfo {volume} {32}},\ \bibinfo {eid} {074001} (\bibinfo {year} {2015})}\BibitemShut {NoStop}%
\bibitem [{\citenamefont {Collaboration}(2014)}]{Acernese15}%
  \BibitemOpen
  \bibfield  {author} {\bibinfo {author} {\bibfnamefont {V.~S.}\ \bibnamefont {Collaboration}},\ }\bibfield  {title} {\bibinfo {title} {Advanced virgo: a second-generation interferometric gravitational wave detector},\ }\href {https://doi.org/10.1088/0264-9381/32/2/024001} {\bibfield  {journal} {\bibinfo  {journal} {Classical and Quantum Gravity}\ }\textbf {\bibinfo {volume} {32}},\ \bibinfo {pages} {024001} (\bibinfo {year} {2014})}\BibitemShut {NoStop}%
\bibitem [{\citenamefont {{KAGRA Scientific Collaboration}}(2019)}]{Akutsu19}%
  \BibitemOpen
  \bibfield  {author} {\bibinfo {author} {\bibnamefont {{KAGRA Scientific Collaboration}}},\ }\bibfield  {title} {\bibinfo {title} {Kagra: 2.5 generation interferometric gravitational wave detector},\ }\href {https://doi.org/10.1038/s41550-018-0658-y} {\bibfield  {journal} {\bibinfo  {journal} {Nature Astronomy 2019 3:1}\ }\textbf {\bibinfo {volume} {3}},\ \bibinfo {pages} {35} (\bibinfo {year} {2019})}\BibitemShut {NoStop}%
\bibitem [{\citenamefont {{Szczepa{\'n}czyk}}\ \emph {et~al.}(2024)\citenamefont {{Szczepa{\'n}czyk}}, \citenamefont {{Zheng}}, \citenamefont {{Antelis}}, \citenamefont {{Benjamin}}, \citenamefont {{Bizouard}}, \citenamefont {{Casallas-Lagos}}, \citenamefont {{Cerd{\'a}-Dur{\'a}n}}, \citenamefont {{Davis}}, \citenamefont {{Gondek-Rosi{\'n}ska}}, \citenamefont {{Klimenko}}, \citenamefont {{Moreno}}, \citenamefont {{Obergaulinger}}, \citenamefont {{Powell}}, \citenamefont {{Ramirez}}, \citenamefont {{Ratto}}, \citenamefont {{Richardson}}, \citenamefont {{Rijal}}, \citenamefont {{Stuver}}, \citenamefont {{Szewczyk}}, \citenamefont {{Vedovato}}, \citenamefont {{Zanolin}}, \citenamefont {{Bartos}}, \citenamefont {{Bhaumik}}, \citenamefont {{Bulik}}, \citenamefont {{Drago}}, \citenamefont {{Font}}, \citenamefont {{De Colle}}, \citenamefont {{Garc{\'\i}a-Bellido}}, \citenamefont {{Gayathri}}, \citenamefont {{Hughey}}, \citenamefont {{Mitselmakher}}, \citenamefont {{Mishra}}, \citenamefont {{Mukherjee}},
  \citenamefont {{Nguyen}}, \citenamefont {{Chan}}, \citenamefont {{Di Palma}}, \citenamefont {{Piotrzkowski}},\ and\ \citenamefont {{Singh}}}]{LVKO3SNsearch23}%
  \BibitemOpen
  \bibfield  {author} {\bibinfo {author} {\bibfnamefont {M.~J.}\ \bibnamefont {{Szczepa{\'n}czyk}}}, \bibinfo {author} {\bibfnamefont {Y.}~\bibnamefont {{Zheng}}}, \bibinfo {author} {\bibfnamefont {J.~M.}\ \bibnamefont {{Antelis}}}, \bibinfo {author} {\bibfnamefont {M.}~\bibnamefont {{Benjamin}}}, \bibinfo {author} {\bibfnamefont {M.-A.}\ \bibnamefont {{Bizouard}}}, \bibinfo {author} {\bibfnamefont {A.}~\bibnamefont {{Casallas-Lagos}}}, \bibinfo {author} {\bibfnamefont {P.}~\bibnamefont {{Cerd{\'a}-Dur{\'a}n}}}, \bibinfo {author} {\bibfnamefont {D.}~\bibnamefont {{Davis}}}, \bibinfo {author} {\bibfnamefont {D.}~\bibnamefont {{Gondek-Rosi{\'n}ska}}}, \bibinfo {author} {\bibfnamefont {S.}~\bibnamefont {{Klimenko}}}, \bibinfo {author} {\bibfnamefont {C.}~\bibnamefont {{Moreno}}}, \bibinfo {author} {\bibfnamefont {M.}~\bibnamefont {{Obergaulinger}}}, \bibinfo {author} {\bibfnamefont {J.}~\bibnamefont {{Powell}}}, \bibinfo {author} {\bibfnamefont {D.}~\bibnamefont {{Ramirez}}}, \bibinfo {author} {\bibfnamefont
  {B.}~\bibnamefont {{Ratto}}}, \bibinfo {author} {\bibfnamefont {C.}~\bibnamefont {{Richardson}}}, \bibinfo {author} {\bibfnamefont {A.}~\bibnamefont {{Rijal}}}, \bibinfo {author} {\bibfnamefont {A.~L.}\ \bibnamefont {{Stuver}}}, \bibinfo {author} {\bibfnamefont {P.}~\bibnamefont {{Szewczyk}}}, \bibinfo {author} {\bibfnamefont {G.}~\bibnamefont {{Vedovato}}}, \bibinfo {author} {\bibfnamefont {M.}~\bibnamefont {{Zanolin}}}, \bibinfo {author} {\bibfnamefont {I.}~\bibnamefont {{Bartos}}}, \bibinfo {author} {\bibfnamefont {S.}~\bibnamefont {{Bhaumik}}}, \bibinfo {author} {\bibfnamefont {T.}~\bibnamefont {{Bulik}}}, \bibinfo {author} {\bibfnamefont {M.}~\bibnamefont {{Drago}}}, \bibinfo {author} {\bibfnamefont {J.~A.}\ \bibnamefont {{Font}}}, \bibinfo {author} {\bibfnamefont {F.}~\bibnamefont {{De Colle}}}, \bibinfo {author} {\bibfnamefont {J.}~\bibnamefont {{Garc{\'\i}a-Bellido}}}, \bibinfo {author} {\bibfnamefont {V.}~\bibnamefont {{Gayathri}}}, \bibinfo {author} {\bibfnamefont {B.}~\bibnamefont {{Hughey}}},
  \bibinfo {author} {\bibfnamefont {G.}~\bibnamefont {{Mitselmakher}}}, \bibinfo {author} {\bibfnamefont {T.}~\bibnamefont {{Mishra}}}, \bibinfo {author} {\bibfnamefont {S.}~\bibnamefont {{Mukherjee}}}, \bibinfo {author} {\bibfnamefont {Q.~L.}\ \bibnamefont {{Nguyen}}}, \bibinfo {author} {\bibfnamefont {M.~L.}\ \bibnamefont {{Chan}}}, \bibinfo {author} {\bibfnamefont {I.}~\bibnamefont {{Di Palma}}}, \bibinfo {author} {\bibfnamefont {B.~J.}\ \bibnamefont {{Piotrzkowski}}},\ and\ \bibinfo {author} {\bibfnamefont {N.}~\bibnamefont {{Singh}}},\ }\bibfield  {title} {\bibinfo {title} {{Optically targeted search for gravitational waves emitted by core-collapse supernovae during the third observing run of Advanced LIGO and Advanced Virgo}},\ }\href {https://doi.org/10.1103/PhysRevD.110.042007} {\bibfield  {journal} {\bibinfo  {journal} {\prd}\ }\textbf {\bibinfo {volume} {110}},\ \bibinfo {eid} {042007} (\bibinfo {year} {2024})}\BibitemShut {NoStop}%
\bibitem [{\citenamefont {{The LIGO Scientific Collaboration}}\ \emph {et~al.}(2025)\citenamefont {{The LIGO Scientific Collaboration}}, \citenamefont {{the Virgo Collaboration}},\ and\ \citenamefont {{the KAGRA Collaboration}}}]{LVKSN2023ixf2024}%
  \BibitemOpen
  \bibfield  {author} {\bibinfo {author} {\bibnamefont {{The LIGO Scientific Collaboration}}}, \bibinfo {author} {\bibnamefont {{the Virgo Collaboration}}},\ and\ \bibinfo {author} {\bibnamefont {{the KAGRA Collaboration}}},\ }\bibfield  {title} {\bibinfo {title} {{Search for Gravitational Waves Emitted from SN 2023ixf}},\ }\href {https://doi.org/10.3847/1538-4357/adc681} {\bibfield  {journal} {\bibinfo  {journal} {\apj}\ }\textbf {\bibinfo {volume} {985}},\ \bibinfo {eid} {183} (\bibinfo {year} {2025})}\BibitemShut {NoStop}%
\bibitem [{\citenamefont {{Maggiore}}\ \emph {et~al.}(2020)\citenamefont {{Maggiore}}, \citenamefont {{Van Den Broeck}}, \citenamefont {{Bartolo}}, \citenamefont {{Belgacem}}, \citenamefont {{Bertacca}}, \citenamefont {{Bizouard}}, \citenamefont {{Branchesi}}, \citenamefont {{Clesse}}, \citenamefont {{Foffa}}, \citenamefont {{Garc{\'\i}a-Bellido}}, \citenamefont {{Grimm}}, \citenamefont {{Harms}}, \citenamefont {{Hinderer}}, \citenamefont {{Matarrese}}, \citenamefont {{Palomba}}, \citenamefont {{Peloso}}, \citenamefont {{Ricciardone}},\ and\ \citenamefont {{Sakellariadou}}}]{Maggiore20}%
  \BibitemOpen
  \bibfield  {author} {\bibinfo {author} {\bibfnamefont {M.}~\bibnamefont {{Maggiore}}}, \bibinfo {author} {\bibfnamefont {C.}~\bibnamefont {{Van Den Broeck}}}, \bibinfo {author} {\bibfnamefont {N.}~\bibnamefont {{Bartolo}}}, \bibinfo {author} {\bibfnamefont {E.}~\bibnamefont {{Belgacem}}}, \bibinfo {author} {\bibfnamefont {D.}~\bibnamefont {{Bertacca}}}, \bibinfo {author} {\bibfnamefont {M.~A.}\ \bibnamefont {{Bizouard}}}, \bibinfo {author} {\bibfnamefont {M.}~\bibnamefont {{Branchesi}}}, \bibinfo {author} {\bibfnamefont {S.}~\bibnamefont {{Clesse}}}, \bibinfo {author} {\bibfnamefont {S.}~\bibnamefont {{Foffa}}}, \bibinfo {author} {\bibfnamefont {J.}~\bibnamefont {{Garc{\'\i}a-Bellido}}}, \bibinfo {author} {\bibfnamefont {S.}~\bibnamefont {{Grimm}}}, \bibinfo {author} {\bibfnamefont {J.}~\bibnamefont {{Harms}}}, \bibinfo {author} {\bibfnamefont {T.}~\bibnamefont {{Hinderer}}}, \bibinfo {author} {\bibfnamefont {S.}~\bibnamefont {{Matarrese}}}, \bibinfo {author} {\bibfnamefont {C.}~\bibnamefont {{Palomba}}},
  \bibinfo {author} {\bibfnamefont {M.}~\bibnamefont {{Peloso}}}, \bibinfo {author} {\bibfnamefont {A.}~\bibnamefont {{Ricciardone}}},\ and\ \bibinfo {author} {\bibfnamefont {M.}~\bibnamefont {{Sakellariadou}}},\ }\bibfield  {title} {\bibinfo {title} {{Science case for the Einstein telescope}},\ }\href {https://doi.org/10.1088/1475-7516/2020/03/050} {\bibfield  {journal} {\bibinfo  {journal} {\jcap}\ }\textbf {\bibinfo {volume} {2020}},\ \bibinfo {eid} {050} (\bibinfo {year} {2020})}\BibitemShut {NoStop}%
\bibitem [{\citenamefont {{Reitze}}\ \emph {et~al.}(2019)\citenamefont {{Reitze}}, \citenamefont {{Adhikari}}, \citenamefont {{Ballmer}}, \citenamefont {{Barish}}, \citenamefont {{Barsotti}}, \citenamefont {{Billingsley}}, \citenamefont {{Brown}}, \citenamefont {{Chen}}, \citenamefont {{Coyne}}, \citenamefont {{Eisenstein}}, \citenamefont {{Evans}}, \citenamefont {{Fritschel}}, \citenamefont {{Hall}}, \citenamefont {{Lazzarini}}, \citenamefont {{Lovelace}}, \citenamefont {{Read}}, \citenamefont {{Sathyaprakash}}, \citenamefont {{Shoemaker}}, \citenamefont {{Smith}}, \citenamefont {{Torrie}}, \citenamefont {{Vitale}}, \citenamefont {{Weiss}}, \citenamefont {{Wipf}},\ and\ \citenamefont {{Zucker}}}]{Reitze19}%
  \BibitemOpen
  \bibfield  {author} {\bibinfo {author} {\bibfnamefont {D.}~\bibnamefont {{Reitze}}}, \bibinfo {author} {\bibfnamefont {R.~X.}\ \bibnamefont {{Adhikari}}}, \bibinfo {author} {\bibfnamefont {S.}~\bibnamefont {{Ballmer}}}, \bibinfo {author} {\bibfnamefont {B.}~\bibnamefont {{Barish}}}, \bibinfo {author} {\bibfnamefont {L.}~\bibnamefont {{Barsotti}}}, \bibinfo {author} {\bibfnamefont {G.}~\bibnamefont {{Billingsley}}}, \bibinfo {author} {\bibfnamefont {D.~A.}\ \bibnamefont {{Brown}}}, \bibinfo {author} {\bibfnamefont {Y.}~\bibnamefont {{Chen}}}, \bibinfo {author} {\bibfnamefont {D.}~\bibnamefont {{Coyne}}}, \bibinfo {author} {\bibfnamefont {R.}~\bibnamefont {{Eisenstein}}}, \bibinfo {author} {\bibfnamefont {M.}~\bibnamefont {{Evans}}}, \bibinfo {author} {\bibfnamefont {P.}~\bibnamefont {{Fritschel}}}, \bibinfo {author} {\bibfnamefont {E.~D.}\ \bibnamefont {{Hall}}}, \bibinfo {author} {\bibfnamefont {A.}~\bibnamefont {{Lazzarini}}}, \bibinfo {author} {\bibfnamefont {G.}~\bibnamefont {{Lovelace}}}, \bibinfo
  {author} {\bibfnamefont {J.}~\bibnamefont {{Read}}}, \bibinfo {author} {\bibfnamefont {B.~S.}\ \bibnamefont {{Sathyaprakash}}}, \bibinfo {author} {\bibfnamefont {D.}~\bibnamefont {{Shoemaker}}}, \bibinfo {author} {\bibfnamefont {J.}~\bibnamefont {{Smith}}}, \bibinfo {author} {\bibfnamefont {C.}~\bibnamefont {{Torrie}}}, \bibinfo {author} {\bibfnamefont {S.}~\bibnamefont {{Vitale}}}, \bibinfo {author} {\bibfnamefont {R.}~\bibnamefont {{Weiss}}}, \bibinfo {author} {\bibfnamefont {C.}~\bibnamefont {{Wipf}}},\ and\ \bibinfo {author} {\bibfnamefont {M.}~\bibnamefont {{Zucker}}},\ }\bibfield  {title} {\bibinfo {title} {{Cosmic Explorer: The U.S. Contribution to Gravitational-Wave Astronomy beyond LIGO}},\ }in\ \href {https://doi.org/10.48550/arXiv.1907.04833} {\emph {\bibinfo {booktitle} {Bulletin of the American Astronomical Society}}},\ Vol.~\bibinfo {volume} {51}\ (\bibinfo {year} {2019})\ p.~\bibinfo {pages} {35}\BibitemShut {NoStop}%
\bibitem [{\citenamefont {{Burrows}}(1988)}]{Burrows88}%
  \BibitemOpen
  \bibfield  {author} {\bibinfo {author} {\bibfnamefont {A.}~\bibnamefont {{Burrows}}},\ }\bibfield  {title} {\bibinfo {title} {{Supernova Neutrinos}},\ }\href {https://doi.org/10.1086/166885} {\bibfield  {journal} {\bibinfo  {journal} {\apj}\ }\textbf {\bibinfo {volume} {334}},\ \bibinfo {pages} {891} (\bibinfo {year} {1988})}\BibitemShut {NoStop}%
\bibitem [{\citenamefont {{M{\"u}ller}}\ \emph {et~al.}(2013)\citenamefont {{M{\"u}ller}}, \citenamefont {{Janka}},\ and\ \citenamefont {{Marek}}}]{Muller13}%
  \BibitemOpen
  \bibfield  {author} {\bibinfo {author} {\bibfnamefont {B.}~\bibnamefont {{M{\"u}ller}}}, \bibinfo {author} {\bibfnamefont {H.-T.}\ \bibnamefont {{Janka}}},\ and\ \bibinfo {author} {\bibfnamefont {A.}~\bibnamefont {{Marek}}},\ }\bibfield  {title} {\bibinfo {title} {{A New Multi-dimensional General Relativistic Neutrino Hydrodynamics Code of Core-collapse Supernovae. III. Gravitational Wave Signals from Supernova Explosion Models}},\ }\href {https://doi.org/10.1088/0004-637X/766/1/43} {\bibfield  {journal} {\bibinfo  {journal} {\apj}\ }\textbf {\bibinfo {volume} {766}},\ \bibinfo {eid} {43} (\bibinfo {year} {2013})}\BibitemShut {NoStop}%
\bibitem [{\citenamefont {{Pan}}\ \emph {et~al.}(2018)\citenamefont {{Pan}}, \citenamefont {{Liebend{\"o}rfer}}, \citenamefont {{Couch}},\ and\ \citenamefont {{Thielemann}}}]{Pan18}%
  \BibitemOpen
  \bibfield  {author} {\bibinfo {author} {\bibfnamefont {K.-C.}\ \bibnamefont {{Pan}}}, \bibinfo {author} {\bibfnamefont {M.}~\bibnamefont {{Liebend{\"o}rfer}}}, \bibinfo {author} {\bibfnamefont {S.~M.}\ \bibnamefont {{Couch}}},\ and\ \bibinfo {author} {\bibfnamefont {F.-K.}\ \bibnamefont {{Thielemann}}},\ }\bibfield  {title} {\bibinfo {title} {{Equation of State Dependent Dynamics and Multi-messenger Signals from Stellar-mass Black Hole Formation}},\ }\href {https://doi.org/10.3847/1538-4357/aab71d} {\bibfield  {journal} {\bibinfo  {journal} {\apj}\ }\textbf {\bibinfo {volume} {857}},\ \bibinfo {eid} {13} (\bibinfo {year} {2018})}\BibitemShut {NoStop}%
\bibitem [{\citenamefont {{Cerd{\'a}-Dur{\'a}n}}\ \emph {et~al.}(2013)\citenamefont {{Cerd{\'a}-Dur{\'a}n}}, \citenamefont {{DeBrye}}, \citenamefont {{Aloy}}, \citenamefont {{Font}},\ and\ \citenamefont {{Obergaulinger}}}]{CerdaDuran13}%
  \BibitemOpen
  \bibfield  {author} {\bibinfo {author} {\bibfnamefont {P.}~\bibnamefont {{Cerd{\'a}-Dur{\'a}n}}}, \bibinfo {author} {\bibfnamefont {N.}~\bibnamefont {{DeBrye}}}, \bibinfo {author} {\bibfnamefont {M.~A.}\ \bibnamefont {{Aloy}}}, \bibinfo {author} {\bibfnamefont {J.~A.}\ \bibnamefont {{Font}}},\ and\ \bibinfo {author} {\bibfnamefont {M.}~\bibnamefont {{Obergaulinger}}},\ }\bibfield  {title} {\bibinfo {title} {{Gravitational Wave Signatures in Black Hole Forming Core Collapse}},\ }\href {https://doi.org/10.1088/2041-8205/779/2/L18} {\bibfield  {journal} {\bibinfo  {journal} {\apjl}\ }\textbf {\bibinfo {volume} {779}},\ \bibinfo {eid} {L18} (\bibinfo {year} {2013})}\BibitemShut {NoStop}%
\bibitem [{\citenamefont {{Torres-Forn{\'e}}}\ \emph {et~al.}(2018)\citenamefont {{Torres-Forn{\'e}}}, \citenamefont {{Cerd{\'a}-Dur{\'a}n}}, \citenamefont {{Passamonti}},\ and\ \citenamefont {{Font}}}]{TorresForne18}%
  \BibitemOpen
  \bibfield  {author} {\bibinfo {author} {\bibfnamefont {A.}~\bibnamefont {{Torres-Forn{\'e}}}}, \bibinfo {author} {\bibfnamefont {P.}~\bibnamefont {{Cerd{\'a}-Dur{\'a}n}}}, \bibinfo {author} {\bibfnamefont {A.}~\bibnamefont {{Passamonti}}},\ and\ \bibinfo {author} {\bibfnamefont {J.~A.}\ \bibnamefont {{Font}}},\ }\bibfield  {title} {\bibinfo {title} {{Towards asteroseismology of core-collapse supernovae with gravitational-wave observations - I. Cowling approximation}},\ }\href {https://doi.org/10.1093/mnras/stx3067} {\bibfield  {journal} {\bibinfo  {journal} {\mnras}\ }\textbf {\bibinfo {volume} {474}},\ \bibinfo {pages} {5272} (\bibinfo {year} {2018})}\BibitemShut {NoStop}%
\bibitem [{\citenamefont {{Andresen}}\ \emph {et~al.}(2017)\citenamefont {{Andresen}}, \citenamefont {{M{\"u}ller}}, \citenamefont {{M{\"u}ller}},\ and\ \citenamefont {{Janka}}}]{Andresen17}%
  \BibitemOpen
  \bibfield  {author} {\bibinfo {author} {\bibfnamefont {H.}~\bibnamefont {{Andresen}}}, \bibinfo {author} {\bibfnamefont {B.}~\bibnamefont {{M{\"u}ller}}}, \bibinfo {author} {\bibfnamefont {E.}~\bibnamefont {{M{\"u}ller}}},\ and\ \bibinfo {author} {\bibfnamefont {H.~T.}\ \bibnamefont {{Janka}}},\ }\bibfield  {title} {\bibinfo {title} {{Gravitational wave signals from 3D neutrino hydrodynamics simulations of core-collapse supernovae}},\ }\href {https://doi.org/10.1093/mnras/stx618} {\bibfield  {journal} {\bibinfo  {journal} {\mnras}\ }\textbf {\bibinfo {volume} {468}},\ \bibinfo {pages} {2032} (\bibinfo {year} {2017})}\BibitemShut {NoStop}%
\bibitem [{\citenamefont {Malik}\ \emph {et~al.}(2018)\citenamefont {Malik}, \citenamefont {Alam}, \citenamefont {Fortin}, \citenamefont {Provid\^encia}, \citenamefont {Agrawal}, \citenamefont {Jha}, \citenamefont {Kumar},\ and\ \citenamefont {Patra}}]{Malik18}%
  \BibitemOpen
  \bibfield  {author} {\bibinfo {author} {\bibfnamefont {T.}~\bibnamefont {Malik}}, \bibinfo {author} {\bibfnamefont {N.}~\bibnamefont {Alam}}, \bibinfo {author} {\bibfnamefont {M.}~\bibnamefont {Fortin}}, \bibinfo {author} {\bibfnamefont {C.}~\bibnamefont {Provid\^encia}}, \bibinfo {author} {\bibfnamefont {B.~K.}\ \bibnamefont {Agrawal}}, \bibinfo {author} {\bibfnamefont {T.~K.}\ \bibnamefont {Jha}}, \bibinfo {author} {\bibfnamefont {B.}~\bibnamefont {Kumar}},\ and\ \bibinfo {author} {\bibfnamefont {S.~K.}\ \bibnamefont {Patra}},\ }\bibfield  {title} {\bibinfo {title} {Gw170817: Constraining the nuclear matter equation of state from the neutron star tidal deformability},\ }\href {https://doi.org/10.1103/PhysRevC.98.035804} {\bibfield  {journal} {\bibinfo  {journal} {Phys. Rev. C}\ }\textbf {\bibinfo {volume} {98}},\ \bibinfo {pages} {035804} (\bibinfo {year} {2018})}\BibitemShut {NoStop}%
\bibitem [{\citenamefont {{Raaijmakers}}\ \emph {et~al.}(2020)\citenamefont {{Raaijmakers}}, \citenamefont {{Greif}}, \citenamefont {{Riley}}, \citenamefont {{Hinderer}}, \citenamefont {{Hebeler}}, \citenamefont {{Schwenk}}, \citenamefont {{Watts}}, \citenamefont {{Nissanke}}, \citenamefont {{Guillot}}, \citenamefont {{Lattimer}},\ and\ \citenamefont {{Ludlam}}}]{Raaijmakers20}%
  \BibitemOpen
  \bibfield  {author} {\bibinfo {author} {\bibfnamefont {G.}~\bibnamefont {{Raaijmakers}}}, \bibinfo {author} {\bibfnamefont {S.~K.}\ \bibnamefont {{Greif}}}, \bibinfo {author} {\bibfnamefont {T.~E.}\ \bibnamefont {{Riley}}}, \bibinfo {author} {\bibfnamefont {T.}~\bibnamefont {{Hinderer}}}, \bibinfo {author} {\bibfnamefont {K.}~\bibnamefont {{Hebeler}}}, \bibinfo {author} {\bibfnamefont {A.}~\bibnamefont {{Schwenk}}}, \bibinfo {author} {\bibfnamefont {A.~L.}\ \bibnamefont {{Watts}}}, \bibinfo {author} {\bibfnamefont {S.}~\bibnamefont {{Nissanke}}}, \bibinfo {author} {\bibfnamefont {S.}~\bibnamefont {{Guillot}}}, \bibinfo {author} {\bibfnamefont {J.~M.}\ \bibnamefont {{Lattimer}}},\ and\ \bibinfo {author} {\bibfnamefont {R.~M.}\ \bibnamefont {{Ludlam}}},\ }\bibfield  {title} {\bibinfo {title} {{Constraining the Dense Matter Equation of State with Joint Analysis of NICER and LIGO/Virgo Measurements}},\ }\href {https://doi.org/10.3847/2041-8213/ab822f} {\bibfield  {journal} {\bibinfo  {journal} {\apjl}\ }\textbf
  {\bibinfo {volume} {893}},\ \bibinfo {eid} {L21} (\bibinfo {year} {2020})}\BibitemShut {NoStop}%
\bibitem [{\citenamefont {Torres-Forn\'e}\ \emph {et~al.}(2019)\citenamefont {Torres-Forn\'e}, \citenamefont {Cerd\'a-Dur\'an}, \citenamefont {Obergaulinger}, \citenamefont {M\"uller},\ and\ \citenamefont {Font}}]{TorresForne19b}%
  \BibitemOpen
  \bibfield  {author} {\bibinfo {author} {\bibfnamefont {A.}~\bibnamefont {Torres-Forn\'e}}, \bibinfo {author} {\bibfnamefont {P.}~\bibnamefont {Cerd\'a-Dur\'an}}, \bibinfo {author} {\bibfnamefont {M.}~\bibnamefont {Obergaulinger}}, \bibinfo {author} {\bibfnamefont {B.}~\bibnamefont {M\"uller}},\ and\ \bibinfo {author} {\bibfnamefont {J.~A.}\ \bibnamefont {Font}},\ }\bibfield  {title} {\bibinfo {title} {Universal relations for gravitational-wave asteroseismology of protoneutron stars},\ }\href {https://doi.org/10.1103/PhysRevLett.123.051102} {\bibfield  {journal} {\bibinfo  {journal} {Phys. Rev. Lett.}\ }\textbf {\bibinfo {volume} {123}},\ \bibinfo {pages} {051102} (\bibinfo {year} {2019})}\BibitemShut {NoStop}%
\bibitem [{\citenamefont {{Bruel}}\ \emph {et~al.}(2023)\citenamefont {{Bruel}}, \citenamefont {{Bizouard}}, \citenamefont {{Obergaulinger}}, \citenamefont {{Maturana-Russel}}, \citenamefont {{Torres-Forn{\'e}}}, \citenamefont {{Cerd{\'a}-Dur{\'a}n}}, \citenamefont {{Christensen}}, \citenamefont {{Font}},\ and\ \citenamefont {{Meyer}}}]{Bruel23}%
  \BibitemOpen
  \bibfield  {author} {\bibinfo {author} {\bibfnamefont {T.}~\bibnamefont {{Bruel}}}, \bibinfo {author} {\bibfnamefont {M.-A.}\ \bibnamefont {{Bizouard}}}, \bibinfo {author} {\bibfnamefont {M.}~\bibnamefont {{Obergaulinger}}}, \bibinfo {author} {\bibfnamefont {P.}~\bibnamefont {{Maturana-Russel}}}, \bibinfo {author} {\bibfnamefont {A.}~\bibnamefont {{Torres-Forn{\'e}}}}, \bibinfo {author} {\bibfnamefont {P.}~\bibnamefont {{Cerd{\'a}-Dur{\'a}n}}}, \bibinfo {author} {\bibfnamefont {N.}~\bibnamefont {{Christensen}}}, \bibinfo {author} {\bibfnamefont {J.~A.}\ \bibnamefont {{Font}}},\ and\ \bibinfo {author} {\bibfnamefont {R.}~\bibnamefont {{Meyer}}},\ }\bibfield  {title} {\bibinfo {title} {{Inference of protoneutron star properties in core-collapse supernovae from a gravitational-wave detector network}},\ }\href {https://doi.org/10.1103/PhysRevD.107.083029} {\bibfield  {journal} {\bibinfo  {journal} {\prd}\ }\textbf {\bibinfo {volume} {107}},\ \bibinfo {eid} {083029} (\bibinfo {year} {2023})}\BibitemShut
  {NoStop}%
\bibitem [{\citenamefont {Rodriguez}\ \emph {et~al.}(2023)\citenamefont {Rodriguez}, \citenamefont {Ranea-Sandoval}, \citenamefont {Chirenti},\ and\ \citenamefont {Radice}}]{Rodriguez23}%
  \BibitemOpen
  \bibfield  {author} {\bibinfo {author} {\bibfnamefont {M.~C.}\ \bibnamefont {Rodriguez}}, \bibinfo {author} {\bibfnamefont {I.~F.}\ \bibnamefont {Ranea-Sandoval}}, \bibinfo {author} {\bibfnamefont {C.}~\bibnamefont {Chirenti}},\ and\ \bibinfo {author} {\bibfnamefont {D.}~\bibnamefont {Radice}},\ }\bibfield  {title} {\bibinfo {title} {{Three approaches for the classification of protoneutron star oscillation modes}},\ }\href {https://doi.org/10.1093/mnras/stad1459} {\bibfield  {journal} {\bibinfo  {journal} {Mon. Not. Roy. Astron. Soc.}\ }\textbf {\bibinfo {volume} {523}},\ \bibinfo {pages} {2236} (\bibinfo {year} {2023})}\BibitemShut {NoStop}%
\bibitem [{\citenamefont {{Shibagaki}}\ \emph {et~al.}(2024)\citenamefont {{Shibagaki}}, \citenamefont {{Kuroda}}, \citenamefont {{Kotake}}, \citenamefont {{Takiwaki}},\ and\ \citenamefont {{Fischer}}}]{Shibagaki24}%
  \BibitemOpen
  \bibfield  {author} {\bibinfo {author} {\bibfnamefont {S.}~\bibnamefont {{Shibagaki}}}, \bibinfo {author} {\bibfnamefont {T.}~\bibnamefont {{Kuroda}}}, \bibinfo {author} {\bibfnamefont {K.}~\bibnamefont {{Kotake}}}, \bibinfo {author} {\bibfnamefont {T.}~\bibnamefont {{Takiwaki}}},\ and\ \bibinfo {author} {\bibfnamefont {T.}~\bibnamefont {{Fischer}}},\ }\bibfield  {title} {\bibinfo {title} {{Three-dimensional GRMHD simulations of rapidly rotating stellar core collapse}},\ }\href {https://doi.org/10.1093/mnras/stae1361} {\bibfield  {journal} {\bibinfo  {journal} {\mnras}\ }\textbf {\bibinfo {volume} {531}},\ \bibinfo {pages} {3732} (\bibinfo {year} {2024})}\BibitemShut {NoStop}%
\bibitem [{\citenamefont {{Powell}}\ \emph {et~al.}(2023)\citenamefont {{Powell}}, \citenamefont {{M{\"u}ller}}, \citenamefont {{Aguilera-Dena}},\ and\ \citenamefont {{Langer}}}]{Powell2023}%
  \BibitemOpen
  \bibfield  {author} {\bibinfo {author} {\bibfnamefont {J.}~\bibnamefont {{Powell}}}, \bibinfo {author} {\bibfnamefont {B.}~\bibnamefont {{M{\"u}ller}}}, \bibinfo {author} {\bibfnamefont {D.~R.}\ \bibnamefont {{Aguilera-Dena}}},\ and\ \bibinfo {author} {\bibfnamefont {N.}~\bibnamefont {{Langer}}},\ }\bibfield  {title} {\bibinfo {title} {{Three dimensional magnetorotational core-collapse supernova explosions of a 39 solar mass progenitor star}},\ }\href {https://doi.org/10.1093/mnras/stad1292} {\bibfield  {journal} {\bibinfo  {journal} {\mnras}\ }\textbf {\bibinfo {volume} {522}},\ \bibinfo {pages} {6070} (\bibinfo {year} {2023})},\ \Eprint {https://arxiv.org/abs/2212.00200} {arXiv:2212.00200 [astro-ph.HE]} \BibitemShut {NoStop}%
\bibitem [{\citenamefont {{Dimmelmeier}}\ \emph {et~al.}(2006)\citenamefont {{Dimmelmeier}}, \citenamefont {{Stergioulas}},\ and\ \citenamefont {{Font}}}]{Dimmelmeier2006}%
  \BibitemOpen
  \bibfield  {author} {\bibinfo {author} {\bibfnamefont {H.}~\bibnamefont {{Dimmelmeier}}}, \bibinfo {author} {\bibfnamefont {N.}~\bibnamefont {{Stergioulas}}},\ and\ \bibinfo {author} {\bibfnamefont {J.~A.}\ \bibnamefont {{Font}}},\ }\bibfield  {title} {\bibinfo {title} {{Non-linear axisymmetric pulsations of rotating relativistic stars in the conformal flatness approximation}},\ }\href {https://doi.org/10.1111/j.1365-2966.2006.10274.x} {\bibfield  {journal} {\bibinfo  {journal} {\mnras}\ }\textbf {\bibinfo {volume} {368}},\ \bibinfo {pages} {1609} (\bibinfo {year} {2006})}\BibitemShut {NoStop}%
\bibitem [{\citenamefont {{Abdikamalov}}\ \emph {et~al.}(2009)\citenamefont {{Abdikamalov}}, \citenamefont {{Dimmelmeier}}, \citenamefont {{Rezzolla}},\ and\ \citenamefont {{Miller}}}]{Abdikamalov09}%
  \BibitemOpen
  \bibfield  {author} {\bibinfo {author} {\bibfnamefont {E.~B.}\ \bibnamefont {{Abdikamalov}}}, \bibinfo {author} {\bibfnamefont {H.}~\bibnamefont {{Dimmelmeier}}}, \bibinfo {author} {\bibfnamefont {L.}~\bibnamefont {{Rezzolla}}},\ and\ \bibinfo {author} {\bibfnamefont {J.~C.}\ \bibnamefont {{Miller}}},\ }\bibfield  {title} {\bibinfo {title} {{Relativistic simulations of the phase-transition-induced collapse of neutron stars}},\ }\href {https://doi.org/10.1111/j.1365-2966.2008.14056.x} {\bibfield  {journal} {\bibinfo  {journal} {\mnras}\ }\textbf {\bibinfo {volume} {392}},\ \bibinfo {pages} {52} (\bibinfo {year} {2009})}\BibitemShut {NoStop}%
\bibitem [{\citenamefont {{Westernacher-Schneider}}\ \emph {et~al.}(2019)\citenamefont {{Westernacher-Schneider}}, \citenamefont {{O'Connor}}, \citenamefont {{O'Sullivan}}, \citenamefont {{Tamborra}}, \citenamefont {{Wu}}, \citenamefont {{Couch}},\ and\ \citenamefont {{Malmenbeck}}}]{Westernacher-Schneider2019}%
  \BibitemOpen
  \bibfield  {author} {\bibinfo {author} {\bibfnamefont {J.~R.}\ \bibnamefont {{Westernacher-Schneider}}}, \bibinfo {author} {\bibfnamefont {E.}~\bibnamefont {{O'Connor}}}, \bibinfo {author} {\bibfnamefont {E.}~\bibnamefont {{O'Sullivan}}}, \bibinfo {author} {\bibfnamefont {I.}~\bibnamefont {{Tamborra}}}, \bibinfo {author} {\bibfnamefont {M.-R.}\ \bibnamefont {{Wu}}}, \bibinfo {author} {\bibfnamefont {S.~M.}\ \bibnamefont {{Couch}}},\ and\ \bibinfo {author} {\bibfnamefont {F.}~\bibnamefont {{Malmenbeck}}},\ }\bibfield  {title} {\bibinfo {title} {{Multimessenger asteroseismology of core-collapse supernovae}},\ }\href {https://doi.org/10.1103/PhysRevD.100.123009} {\bibfield  {journal} {\bibinfo  {journal} {\prd}\ }\textbf {\bibinfo {volume} {100}},\ \bibinfo {eid} {123009} (\bibinfo {year} {2019})}\BibitemShut {NoStop}%
\bibitem [{\citenamefont {{Obergaulinger}}\ \emph {et~al.}(2018)\citenamefont {{Obergaulinger}}, \citenamefont {{Just}},\ and\ \citenamefont {{Aloy}}}]{Obergaulinger_2018JPhG...45h4001}%
  \BibitemOpen
  \bibfield  {author} {\bibinfo {author} {\bibfnamefont {M.}~\bibnamefont {{Obergaulinger}}}, \bibinfo {author} {\bibfnamefont {O.}~\bibnamefont {{Just}}},\ and\ \bibinfo {author} {\bibfnamefont {M.~A.}\ \bibnamefont {{Aloy}}},\ }\bibfield  {title} {\bibinfo {title} {{Core collapse with magnetic fields and rotation}},\ }\href {https://doi.org/10.1088/1361-6471/aac982} {\bibfield  {journal} {\bibinfo  {journal} {Journal of Physics G Nuclear Physics}\ }\textbf {\bibinfo {volume} {45}},\ \bibinfo {pages} {084001} (\bibinfo {year} {2018})}\BibitemShut {NoStop}%
\bibitem [{\citenamefont {{Just}}\ \emph {et~al.}(2015)\citenamefont {{Just}}, \citenamefont {{Obergaulinger}},\ and\ \citenamefont {{Janka}}}]{Just15}%
  \BibitemOpen
  \bibfield  {author} {\bibinfo {author} {\bibfnamefont {O.}~\bibnamefont {{Just}}}, \bibinfo {author} {\bibfnamefont {M.}~\bibnamefont {{Obergaulinger}}},\ and\ \bibinfo {author} {\bibfnamefont {H.~T.}\ \bibnamefont {{Janka}}},\ }\bibfield  {title} {\bibinfo {title} {{A new multidimensional, energy-dependent two-moment transport code for neutrino-hydrodynamics}},\ }\href {https://doi.org/10.1093/mnras/stv1892} {\bibfield  {journal} {\bibinfo  {journal} {\mnras}\ }\textbf {\bibinfo {volume} {453}},\ \bibinfo {pages} {3386} (\bibinfo {year} {2015})}\BibitemShut {NoStop}%
\bibitem [{\citenamefont {{Obergaulinger}}\ and\ \citenamefont {{Aloy}}(2022)}]{Obergaulinger_2022MNRAS.512.2489}%
  \BibitemOpen
  \bibfield  {author} {\bibinfo {author} {\bibfnamefont {M.}~\bibnamefont {{Obergaulinger}}}\ and\ \bibinfo {author} {\bibfnamefont {M.~{\'A}.}\ \bibnamefont {{Aloy}}},\ }\bibfield  {title} {\bibinfo {title} {{Magnetorotational core collapse of possible gamma-ray burst progenitors - IV. A wider range of progenitors}},\ }\href {https://doi.org/10.1093/mnras/stac613} {\bibfield  {journal} {\bibinfo  {journal} {\mnras}\ }\textbf {\bibinfo {volume} {512}},\ \bibinfo {pages} {2489} (\bibinfo {year} {2022})}\BibitemShut {NoStop}%
\bibitem [{\citenamefont {{Marek}}\ \emph {et~al.}(2006)\citenamefont {{Marek}}, \citenamefont {{Dimmelmeier}}, \citenamefont {{Janka}}, \citenamefont {{M{\"u}ller}},\ and\ \citenamefont {{Buras}}}]{Marek06}%
  \BibitemOpen
  \bibfield  {author} {\bibinfo {author} {\bibfnamefont {A.}~\bibnamefont {{Marek}}}, \bibinfo {author} {\bibfnamefont {H.}~\bibnamefont {{Dimmelmeier}}}, \bibinfo {author} {\bibfnamefont {H.~T.}\ \bibnamefont {{Janka}}}, \bibinfo {author} {\bibfnamefont {E.}~\bibnamefont {{M{\"u}ller}}},\ and\ \bibinfo {author} {\bibfnamefont {R.}~\bibnamefont {{Buras}}},\ }\bibfield  {title} {\bibinfo {title} {{Exploring the relativistic regime with Newtonian hydrodynamics: an improved effective gravitational potential for supernova simulations}},\ }\href {https://doi.org/10.1051/0004-6361:20052840} {\bibfield  {journal} {\bibinfo  {journal} {\aap}\ }\textbf {\bibinfo {volume} {445}},\ \bibinfo {pages} {273} (\bibinfo {year} {2006})}\BibitemShut {NoStop}%
\bibitem [{\citenamefont {{Steiner}}\ \emph {et~al.}(2013)\citenamefont {{Steiner}}, \citenamefont {{Lattimer}},\ and\ \citenamefont {{Brown}}}]{Steiner13}%
  \BibitemOpen
  \bibfield  {author} {\bibinfo {author} {\bibfnamefont {A.~W.}\ \bibnamefont {{Steiner}}}, \bibinfo {author} {\bibfnamefont {J.~M.}\ \bibnamefont {{Lattimer}}},\ and\ \bibinfo {author} {\bibfnamefont {E.~F.}\ \bibnamefont {{Brown}}},\ }\bibfield  {title} {\bibinfo {title} {{The Neutron Star Mass-Radius Relation and the Equation of State of Dense Matter}},\ }\href {https://doi.org/10.1088/2041-8205/765/1/L5} {\bibfield  {journal} {\bibinfo  {journal} {\apjl}\ }\textbf {\bibinfo {volume} {765}},\ \bibinfo {eid} {L5} (\bibinfo {year} {2013})}\BibitemShut {NoStop}%
\bibitem [{\citenamefont {{Obergaulinger}}\ \emph {et~al.}(2006)\citenamefont {{Obergaulinger}}, \citenamefont {{Aloy}},\ and\ \citenamefont {{M{\"u}ller}}}]{Obergaulinger_2006A&A...450.1107}%
  \BibitemOpen
  \bibfield  {author} {\bibinfo {author} {\bibfnamefont {M.}~\bibnamefont {{Obergaulinger}}}, \bibinfo {author} {\bibfnamefont {M.~A.}\ \bibnamefont {{Aloy}}},\ and\ \bibinfo {author} {\bibfnamefont {E.}~\bibnamefont {{M{\"u}ller}}},\ }\bibfield  {title} {\bibinfo {title} {{Axisymmetric simulations of magneto-rotational core collapse: dynamics and gravitational wave signal}},\ }\href {https://doi.org/10.1051/0004-6361:20054306} {\bibfield  {journal} {\bibinfo  {journal} {\aap}\ }\textbf {\bibinfo {volume} {450}},\ \bibinfo {pages} {1107} (\bibinfo {year} {2006})}\BibitemShut {NoStop}%
\bibitem [{\citenamefont {{Aguilera-Dena}}\ \emph {et~al.}(2018)\citenamefont {{Aguilera-Dena}}, \citenamefont {{Langer}}, \citenamefont {{Moriya}},\ and\ \citenamefont {{Schootemeijer}}}]{AguileraDena18}%
  \BibitemOpen
  \bibfield  {author} {\bibinfo {author} {\bibfnamefont {D.~R.}\ \bibnamefont {{Aguilera-Dena}}}, \bibinfo {author} {\bibfnamefont {N.}~\bibnamefont {{Langer}}}, \bibinfo {author} {\bibfnamefont {T.~J.}\ \bibnamefont {{Moriya}}},\ and\ \bibinfo {author} {\bibfnamefont {A.}~\bibnamefont {{Schootemeijer}}},\ }\bibfield  {title} {\bibinfo {title} {{Related Progenitor Models for Long-duration Gamma-Ray Bursts and Type Ic Superluminous Supernovae}},\ }\href {https://doi.org/10.3847/1538-4357/aabfc1} {\bibfield  {journal} {\bibinfo  {journal} {\apj}\ }\textbf {\bibinfo {volume} {858}},\ \bibinfo {eid} {115} (\bibinfo {year} {2018})}\BibitemShut {NoStop}%
\bibitem [{\citenamefont {{Woosley}}\ and\ \citenamefont {{Heger}}(2006)}]{Woosley_2006ApJ...637..914}%
  \BibitemOpen
  \bibfield  {author} {\bibinfo {author} {\bibfnamefont {S.~E.}\ \bibnamefont {{Woosley}}}\ and\ \bibinfo {author} {\bibfnamefont {A.}~\bibnamefont {{Heger}}},\ }\bibfield  {title} {\bibinfo {title} {{The Progenitor Stars of Gamma-Ray Bursts}},\ }\href {https://doi.org/10.1086/498500} {\bibfield  {journal} {\bibinfo  {journal} {\apj}\ }\textbf {\bibinfo {volume} {637}},\ \bibinfo {pages} {914} (\bibinfo {year} {2006})}\BibitemShut {NoStop}%
\bibitem [{\citenamefont {{Griffiths}}\ \emph {et~al.}(2022)\citenamefont {{Griffiths}}, \citenamefont {{Eggenberger}}, \citenamefont {{Meynet}}, \citenamefont {{Moyano}},\ and\ \citenamefont {{Aloy}}}]{Griffiths_2022A&A...665A.147}%
  \BibitemOpen
  \bibfield  {author} {\bibinfo {author} {\bibfnamefont {A.}~\bibnamefont {{Griffiths}}}, \bibinfo {author} {\bibfnamefont {P.}~\bibnamefont {{Eggenberger}}}, \bibinfo {author} {\bibfnamefont {G.}~\bibnamefont {{Meynet}}}, \bibinfo {author} {\bibfnamefont {F.}~\bibnamefont {{Moyano}}},\ and\ \bibinfo {author} {\bibfnamefont {M.-{\'A}.}\ \bibnamefont {{Aloy}}},\ }\bibfield  {title} {\bibinfo {title} {{The magneto-rotational instability in massive stars}},\ }\href {https://doi.org/10.1051/0004-6361/202243599} {\bibfield  {journal} {\bibinfo  {journal} {\aap}\ }\textbf {\bibinfo {volume} {665}},\ \bibinfo {eid} {A147} (\bibinfo {year} {2022})}\BibitemShut {NoStop}%
\bibitem [{\citenamefont {{Torres-Forn{\'e}}}\ \emph {et~al.}(2019)\citenamefont {{Torres-Forn{\'e}}}, \citenamefont {{Cerd{\'a}-Dur{\'a}n}}, \citenamefont {{Passamonti}}, \citenamefont {{Obergaulinger}},\ and\ \citenamefont {{Font}}}]{TorresForne19}%
  \BibitemOpen
  \bibfield  {author} {\bibinfo {author} {\bibfnamefont {A.}~\bibnamefont {{Torres-Forn{\'e}}}}, \bibinfo {author} {\bibfnamefont {P.}~\bibnamefont {{Cerd{\'a}-Dur{\'a}n}}}, \bibinfo {author} {\bibfnamefont {A.}~\bibnamefont {{Passamonti}}}, \bibinfo {author} {\bibfnamefont {M.}~\bibnamefont {{Obergaulinger}}},\ and\ \bibinfo {author} {\bibfnamefont {J.~A.}\ \bibnamefont {{Font}}},\ }\bibfield  {title} {\bibinfo {title} {{Towards asteroseismology of core-collapse supernovae with gravitational wave observations - II. Inclusion of space-time perturbations}},\ }\href {https://doi.org/10.1093/mnras/sty2854} {\bibfield  {journal} {\bibinfo  {journal} {\mnras}\ }\textbf {\bibinfo {volume} {482}},\ \bibinfo {pages} {3967} (\bibinfo {year} {2019})}\BibitemShut {NoStop}%
\bibitem [{\citenamefont {{Abdikamalov}}\ \emph {et~al.}(2014)\citenamefont {{Abdikamalov}}, \citenamefont {{Gossan}}, \citenamefont {{DeMaio}},\ and\ \citenamefont {{Ott}}}]{Abdikamalov14}%
  \BibitemOpen
  \bibfield  {author} {\bibinfo {author} {\bibfnamefont {E.}~\bibnamefont {{Abdikamalov}}}, \bibinfo {author} {\bibfnamefont {S.}~\bibnamefont {{Gossan}}}, \bibinfo {author} {\bibfnamefont {A.~M.}\ \bibnamefont {{DeMaio}}},\ and\ \bibinfo {author} {\bibfnamefont {C.~D.}\ \bibnamefont {{Ott}}},\ }\bibfield  {title} {\bibinfo {title} {{Measuring the angular momentum distribution in core-collapse supernova progenitors with gravitational waves}},\ }\href {https://doi.org/10.1103/PhysRevD.90.044001} {\bibfield  {journal} {\bibinfo  {journal} {\prd}\ }\textbf {\bibinfo {volume} {90}},\ \bibinfo {eid} {044001} (\bibinfo {year} {2014})}\BibitemShut {NoStop}%
\bibitem [{\citenamefont {{Richers}}\ \emph {et~al.}(2017)\citenamefont {{Richers}}, \citenamefont {{Ott}}, \citenamefont {{Abdikamalov}}, \citenamefont {{O'Connor}},\ and\ \citenamefont {{Sullivan}}}]{Richers17}%
  \BibitemOpen
  \bibfield  {author} {\bibinfo {author} {\bibfnamefont {S.}~\bibnamefont {{Richers}}}, \bibinfo {author} {\bibfnamefont {C.~D.}\ \bibnamefont {{Ott}}}, \bibinfo {author} {\bibfnamefont {E.}~\bibnamefont {{Abdikamalov}}}, \bibinfo {author} {\bibfnamefont {E.}~\bibnamefont {{O'Connor}}},\ and\ \bibinfo {author} {\bibfnamefont {C.}~\bibnamefont {{Sullivan}}},\ }\bibfield  {title} {\bibinfo {title} {{Equation of state effects on gravitational waves from rotating core collapse}},\ }\href {https://doi.org/10.1103/PhysRevD.95.063019} {\bibfield  {journal} {\bibinfo  {journal} {\prd}\ }\textbf {\bibinfo {volume} {95}},\ \bibinfo {eid} {063019} (\bibinfo {year} {2017})}\BibitemShut {NoStop}%
\bibitem [{\citenamefont {{Zha}}(2024)}]{Zha_2024PhRvD.110h3034}%
  \BibitemOpen
  \bibfield  {author} {\bibinfo {author} {\bibfnamefont {S.}~\bibnamefont {{Zha}}},\ }\bibfield  {title} {\bibinfo {title} {{Proper way to spatially decompose the gravitational-wave origin in stellar collapse simulations}},\ }\href {https://doi.org/10.1103/PhysRevD.110.083034} {\bibfield  {journal} {\bibinfo  {journal} {\prd}\ }\textbf {\bibinfo {volume} {110}},\ \bibinfo {eid} {083034} (\bibinfo {year} {2024})}\BibitemShut {NoStop}%
\bibitem [{\citenamefont {{Ertl}}\ \emph {et~al.}(2016)\citenamefont {{Ertl}}, \citenamefont {{Janka}}, \citenamefont {{Woosley}}, \citenamefont {{Sukhbold}},\ and\ \citenamefont {{Ugliano}}}]{Ertl16}%
  \BibitemOpen
  \bibfield  {author} {\bibinfo {author} {\bibfnamefont {T.}~\bibnamefont {{Ertl}}}, \bibinfo {author} {\bibfnamefont {H.~T.}\ \bibnamefont {{Janka}}}, \bibinfo {author} {\bibfnamefont {S.~E.}\ \bibnamefont {{Woosley}}}, \bibinfo {author} {\bibfnamefont {T.}~\bibnamefont {{Sukhbold}}},\ and\ \bibinfo {author} {\bibfnamefont {M.}~\bibnamefont {{Ugliano}}},\ }\bibfield  {title} {\bibinfo {title} {{A Two-parameter Criterion for Classifying the Explodability of Massive Stars by the Neutrino-driven Mechanism}},\ }\href {https://doi.org/10.3847/0004-637X/818/2/124} {\bibfield  {journal} {\bibinfo  {journal} {\apj}\ }\textbf {\bibinfo {volume} {818}},\ \bibinfo {eid} {124} (\bibinfo {year} {2016})}\BibitemShut {NoStop}%
\bibitem [{\citenamefont {{Moore}}\ \emph {et~al.}(2015)\citenamefont {{Moore}}, \citenamefont {{Cole}},\ and\ \citenamefont {{Berry}}}]{Moore2015}%
  \BibitemOpen
  \bibfield  {author} {\bibinfo {author} {\bibfnamefont {C.~J.}\ \bibnamefont {{Moore}}}, \bibinfo {author} {\bibfnamefont {R.~H.}\ \bibnamefont {{Cole}}},\ and\ \bibinfo {author} {\bibfnamefont {C.~P.~L.}\ \bibnamefont {{Berry}}},\ }\bibfield  {title} {\bibinfo {title} {{Gravitational-wave sensitivity curves}},\ }\href {https://doi.org/10.1088/0264-9381/32/1/015014} {\bibfield  {journal} {\bibinfo  {journal} {Classical and Quantum Gravity}\ }\textbf {\bibinfo {volume} {32}},\ \bibinfo {eid} {015014} (\bibinfo {year} {2015})},\ \Eprint {https://arxiv.org/abs/1408.0740} {arXiv:1408.0740 [gr-qc]} \BibitemShut {NoStop}%
\bibitem [{\citenamefont {Harris}(1978)}]{Harris_1978}%
  \BibitemOpen
  \bibfield  {author} {\bibinfo {author} {\bibfnamefont {F.}~\bibnamefont {Harris}},\ }\bibfield  {title} {\bibinfo {title} {On the use of windows for harmonic analysis with the discrete fourier transform},\ }\href {https://doi.org/10.1109/PROC.1978.10837} {\bibfield  {journal} {\bibinfo  {journal} {Proceedings of the IEEE}\ }\textbf {\bibinfo {volume} {66}},\ \bibinfo {pages} {51} (\bibinfo {year} {1978})}\BibitemShut {NoStop}%
\bibitem [{\citenamefont {{KAGRA Collaboration}}\ and\ \citenamefont {{VIRGO Collaboration}}(2020)}]{LVK20}%
  \BibitemOpen
  \bibfield  {author} {\bibinfo {author} {\bibfnamefont {L.~S.~C.}\ \bibnamefont {{KAGRA Collaboration}}}\ and\ \bibinfo {author} {\bibnamefont {{VIRGO Collaboration}}},\ }\bibfield  {title} {\bibinfo {title} {{Prospects for observing and localizing gravitational-wave transients with Advanced LIGO, Advanced Virgo and KAGRA}},\ }\href {https://doi.org/10.1007/s41114-020-00026-9} {\bibfield  {journal} {\bibinfo  {journal} {Living Reviews in Relativity}\ }\textbf {\bibinfo {volume} {23}},\ \bibinfo {eid} {3} (\bibinfo {year} {2020})}\BibitemShut {NoStop}%
\bibitem [{\citenamefont {{KAGRA Collaboration}}\ and\ \citenamefont {{VIRGO Collaboration}}(2022)}]{LVKASDs}%
  \BibitemOpen
  \bibfield  {author} {\bibinfo {author} {\bibfnamefont {L.~S.~C.}\ \bibnamefont {{KAGRA Collaboration}}}\ and\ \bibinfo {author} {\bibnamefont {{VIRGO Collaboration}}},\ }\href@noop {} {\bibinfo {title} {{Noise curves used for Simulations in the update of the Observing Scenarios Paper}}},\ \bibinfo {howpublished} {\url{https://dcc.ligo.org/LIGO-T2000012/public}} (\bibinfo {year} {2022}),\ \bibinfo {note} {online; accessed January 2025}\BibitemShut {NoStop}%
\bibitem [{\citenamefont {{Hild}}\ and\ \citenamefont {{others}}(2011)}]{Hild11}%
  \BibitemOpen
  \bibfield  {author} {\bibinfo {author} {\bibfnamefont {S.}~\bibnamefont {{Hild}}}\ and\ \bibinfo {author} {\bibnamefont {{others}}},\ }\bibfield  {title} {\bibinfo {title} {{Sensitivity studies for third-generation gravitational wave observatories}},\ }\href {https://doi.org/10.1088/0264-9381/28/9/094013} {\bibfield  {journal} {\bibinfo  {journal} {Classical and Quantum Gravity}\ }\textbf {\bibinfo {volume} {28}},\ \bibinfo {eid} {094013} (\bibinfo {year} {2011})},\ \Eprint {https://arxiv.org/abs/1012.0908} {arXiv:1012.0908 [gr-qc]} \BibitemShut {NoStop}%
\bibitem [{\citenamefont {{ET design team}}(2018)}]{ETASD}%
  \BibitemOpen
  \bibfield  {author} {\bibinfo {author} {\bibnamefont {{ET design team}}},\ }\href@noop {} {\bibinfo {title} {{ET-D sensitivity curve}}},\ \bibinfo {howpublished} {\url{https://apps.et-gw.eu/tds/?content=3&r=14065}} (\bibinfo {year} {2018}),\ \bibinfo {note} {online; accessed January 2025}\BibitemShut {NoStop}%
\bibitem [{\citenamefont {{Srivastava}}\ \emph {et~al.}(2022)\citenamefont {{Srivastava}}, \citenamefont {{Davis}}, \citenamefont {{Kuns}}, \citenamefont {{Landry}}, \citenamefont {{Ballmer}}, \citenamefont {{Evans}}, \citenamefont {{Hall}}, \citenamefont {{Read}},\ and\ \citenamefont {{Sathyaprakash}}}]{Srivastava22}%
  \BibitemOpen
  \bibfield  {author} {\bibinfo {author} {\bibfnamefont {V.}~\bibnamefont {{Srivastava}}}, \bibinfo {author} {\bibfnamefont {D.}~\bibnamefont {{Davis}}}, \bibinfo {author} {\bibfnamefont {K.}~\bibnamefont {{Kuns}}}, \bibinfo {author} {\bibfnamefont {P.}~\bibnamefont {{Landry}}}, \bibinfo {author} {\bibfnamefont {S.}~\bibnamefont {{Ballmer}}}, \bibinfo {author} {\bibfnamefont {M.}~\bibnamefont {{Evans}}}, \bibinfo {author} {\bibfnamefont {E.~D.}\ \bibnamefont {{Hall}}}, \bibinfo {author} {\bibfnamefont {J.}~\bibnamefont {{Read}}},\ and\ \bibinfo {author} {\bibfnamefont {B.~S.}\ \bibnamefont {{Sathyaprakash}}},\ }\bibfield  {title} {\bibinfo {title} {{Science-driven Tunable Design of Cosmic Explorer Detectors}},\ }\href {https://doi.org/10.3847/1538-4357/ac5f04} {\bibfield  {journal} {\bibinfo  {journal} {\apj}\ }\textbf {\bibinfo {volume} {931}},\ \bibinfo {eid} {22} (\bibinfo {year} {2022})}\BibitemShut {NoStop}%
\bibitem [{\citenamefont {{{Kuns}, Kevin and {Fulda}, Paul and {Barsotti}, Lisa and {Evans}, Matthew}}(2023)}]{CEASD}%
  \BibitemOpen
  \bibfield  {author} {\bibinfo {author} {\bibnamefont {{{Kuns}, Kevin and {Fulda}, Paul and {Barsotti}, Lisa and {Evans}, Matthew}}},\ }\href@noop {} {\bibinfo {title} {{Cosmic Explorer Document T2000017-v8}}},\ \bibinfo {howpublished} {\url{https://dcc.cosmicexplorer.org/CE-T2000017/public}} (\bibinfo {year} {2023}),\ \bibinfo {note} {online; accessed January 2025}\BibitemShut {NoStop}%
\bibitem [{\citenamefont {{Powell}}\ and\ \citenamefont {{M{\"u}ller}}(2024)}]{Powell2024}%
  \BibitemOpen
  \bibfield  {author} {\bibinfo {author} {\bibfnamefont {J.}~\bibnamefont {{Powell}}}\ and\ \bibinfo {author} {\bibfnamefont {B.}~\bibnamefont {{M{\"u}ller}}},\ }\bibfield  {title} {\bibinfo {title} {{The gravitational-wave emission from the explosion of a 15 solar mass star with rotation and magnetic fields}},\ }\href {https://doi.org/10.1093/mnras/stae1731} {\bibfield  {journal} {\bibinfo  {journal} {\mnras}\ }\textbf {\bibinfo {volume} {532}},\ \bibinfo {pages} {4326} (\bibinfo {year} {2024})},\ \Eprint {https://arxiv.org/abs/2406.09691} {arXiv:2406.09691 [astro-ph.HE]} \BibitemShut {NoStop}%
\bibitem [{\citenamefont {{Karachentsev}}\ and\ \citenamefont {{Kashibadze}}(2006)}]{Karachentsev2006}%
  \BibitemOpen
  \bibfield  {author} {\bibinfo {author} {\bibfnamefont {I.~D.}\ \bibnamefont {{Karachentsev}}}\ and\ \bibinfo {author} {\bibfnamefont {O.~G.}\ \bibnamefont {{Kashibadze}}},\ }\bibfield  {title} {\bibinfo {title} {{Masses of the local group and of the M81 group estimated from distortions in the local velocity field}},\ }\href {https://doi.org/10.1007/s10511-006-0002-6} {\bibfield  {journal} {\bibinfo  {journal} {Astrophysics}\ }\textbf {\bibinfo {volume} {49}},\ \bibinfo {pages} {3} (\bibinfo {year} {2006})}\BibitemShut {NoStop}%
\bibitem [{\citenamefont {{Suvorova}}\ \emph {et~al.}(2019)\citenamefont {{Suvorova}}, \citenamefont {{Powell}},\ and\ \citenamefont {{Melatos}}}]{Suvorova19}%
  \BibitemOpen
  \bibfield  {author} {\bibinfo {author} {\bibfnamefont {S.}~\bibnamefont {{Suvorova}}}, \bibinfo {author} {\bibfnamefont {J.}~\bibnamefont {{Powell}}},\ and\ \bibinfo {author} {\bibfnamefont {A.}~\bibnamefont {{Melatos}}},\ }\bibfield  {title} {\bibinfo {title} {{Reconstructing gravitational wave core-collapse supernova signals with dynamic time warping}},\ }\href {https://doi.org/10.1103/PhysRevD.99.123012} {\bibfield  {journal} {\bibinfo  {journal} {\prd}\ }\textbf {\bibinfo {volume} {99}},\ \bibinfo {eid} {123012} (\bibinfo {year} {2019})}\BibitemShut {NoStop}%
\bibitem [{\citenamefont {{Murphy}}\ \emph {et~al.}(2024)\citenamefont {{Murphy}}, \citenamefont {{Casallas-Lagos}}, \citenamefont {{Mezzacappa}}, \citenamefont {{Zanolin}}, \citenamefont {{Landfield}}, \citenamefont {{Lentz}}, \citenamefont {{Marronetti}}, \citenamefont {{Antelis}},\ and\ \citenamefont {{Moreno}}}]{Murphy24}%
  \BibitemOpen
  \bibfield  {author} {\bibinfo {author} {\bibfnamefont {R.~D.}\ \bibnamefont {{Murphy}}}, \bibinfo {author} {\bibfnamefont {A.}~\bibnamefont {{Casallas-Lagos}}}, \bibinfo {author} {\bibfnamefont {A.}~\bibnamefont {{Mezzacappa}}}, \bibinfo {author} {\bibfnamefont {M.}~\bibnamefont {{Zanolin}}}, \bibinfo {author} {\bibfnamefont {R.~E.}\ \bibnamefont {{Landfield}}}, \bibinfo {author} {\bibfnamefont {E.~J.}\ \bibnamefont {{Lentz}}}, \bibinfo {author} {\bibfnamefont {P.}~\bibnamefont {{Marronetti}}}, \bibinfo {author} {\bibfnamefont {J.~M.}\ \bibnamefont {{Antelis}}},\ and\ \bibinfo {author} {\bibfnamefont {C.}~\bibnamefont {{Moreno}}},\ }\bibfield  {title} {\bibinfo {title} {{Dependence of the reconstructed core-collapse supernova gravitational wave high-frequency feature on the nuclear equation of state in real interferometric data}},\ }\href {https://doi.org/10.1103/PhysRevD.110.083006} {\bibfield  {journal} {\bibinfo  {journal} {\prd}\ }\textbf {\bibinfo {volume} {110}},\ \bibinfo {eid} {083006} (\bibinfo
  {year} {2024})}\BibitemShut {NoStop}%
\bibitem [{\citenamefont {{Holgado}}\ \emph {et~al.}(2022)\citenamefont {{Holgado}}, \citenamefont {{Sim{\'o}n-D{\'\i}az}}, \citenamefont {{Herrero}},\ and\ \citenamefont {{Barb{\'a}}}}]{Holgado22}%
  \BibitemOpen
  \bibfield  {author} {\bibinfo {author} {\bibfnamefont {G.}~\bibnamefont {{Holgado}}}, \bibinfo {author} {\bibfnamefont {S.}~\bibnamefont {{Sim{\'o}n-D{\'\i}az}}}, \bibinfo {author} {\bibfnamefont {A.}~\bibnamefont {{Herrero}}},\ and\ \bibinfo {author} {\bibfnamefont {R.~H.}\ \bibnamefont {{Barb{\'a}}}},\ }\bibfield  {title} {\bibinfo {title} {{The IACOB project. VII. The rotational properties of Galactic massive O-type stars revisited}},\ }\href {https://doi.org/10.1051/0004-6361/202243851} {\bibfield  {journal} {\bibinfo  {journal} {\aap}\ }\textbf {\bibinfo {volume} {665}},\ \bibinfo {eid} {A150} (\bibinfo {year} {2022})}\BibitemShut {NoStop}%
\bibitem [{\citenamefont {{Ando}}(2004)}]{Ando_2004ApJ...607...20}%
  \BibitemOpen
  \bibfield  {author} {\bibinfo {author} {\bibfnamefont {S.}~\bibnamefont {{Ando}}},\ }\bibfield  {title} {\bibinfo {title} {{Cosmic Star Formation History and the Future Observation of Supernova Relic Neutrinos}},\ }\href {https://doi.org/10.1086/383467} {\bibfield  {journal} {\bibinfo  {journal} {\apj}\ }\textbf {\bibinfo {volume} {607}},\ \bibinfo {pages} {20} (\bibinfo {year} {2004})}\BibitemShut {NoStop}%
\bibitem [{\citenamefont {{Botticella}}\ \emph {et~al.}(2012)\citenamefont {{Botticella}}, \citenamefont {{Riello}}, \citenamefont {{Cappellaro}}, \citenamefont {{Benetti}}, \citenamefont {{Altavilla}}, \citenamefont {{Pastorello}}, \citenamefont {{Pignata}}, \citenamefont {{Taubenberger}},\ and\ \citenamefont {{Valenti}}}]{Botticella_2012A&A...537A.132}%
  \BibitemOpen
  \bibfield  {author} {\bibinfo {author} {\bibfnamefont {M.~T.}\ \bibnamefont {{Botticella}}}, \bibinfo {author} {\bibfnamefont {M.}~\bibnamefont {{Riello}}}, \bibinfo {author} {\bibfnamefont {E.}~\bibnamefont {{Cappellaro}}}, \bibinfo {author} {\bibfnamefont {S.}~\bibnamefont {{Benetti}}}, \bibinfo {author} {\bibfnamefont {G.}~\bibnamefont {{Altavilla}}}, \bibinfo {author} {\bibfnamefont {A.}~\bibnamefont {{Pastorello}}}, \bibinfo {author} {\bibfnamefont {G.}~\bibnamefont {{Pignata}}}, \bibinfo {author} {\bibfnamefont {S.}~\bibnamefont {{Taubenberger}}},\ and\ \bibinfo {author} {\bibfnamefont {S.}~\bibnamefont {{Valenti}}},\ }\bibfield  {title} {\bibinfo {title} {{Supernova rates from the SUDARE survey: the local rate of core-collapse supernovae}},\ }\href {https://doi.org/10.1051/0004-6361/201117324} {\bibfield  {journal} {\bibinfo  {journal} {\aap}\ }\textbf {\bibinfo {volume} {537}},\ \bibinfo {eid} {A132} (\bibinfo {year} {2012})}\BibitemShut {NoStop}%
\bibitem [{\citenamefont {{Horiuchi}}\ \emph {et~al.}(2013)\citenamefont {{Horiuchi}}, \citenamefont {{Beacom}}, \citenamefont {{Bothwell}}, \citenamefont {{Thompson}},\ and\ \citenamefont {{Szczygie{\l}}}}]{Horiuchi_2013ApJ...769..113}%
  \BibitemOpen
  \bibfield  {author} {\bibinfo {author} {\bibfnamefont {S.}~\bibnamefont {{Horiuchi}}}, \bibinfo {author} {\bibfnamefont {J.~F.}\ \bibnamefont {{Beacom}}}, \bibinfo {author} {\bibfnamefont {M.~S.}\ \bibnamefont {{Bothwell}}}, \bibinfo {author} {\bibfnamefont {T.~A.}\ \bibnamefont {{Thompson}}},\ and\ \bibinfo {author} {\bibfnamefont {D.~M.}\ \bibnamefont {{Szczygie{\l}}}},\ }\bibfield  {title} {\bibinfo {title} {{The Cosmic Core-collapse Supernova Rate Does Not Match the Massive-star Formation Rate}},\ }\href {https://doi.org/10.1088/0004-637X/769/2/113} {\bibfield  {journal} {\bibinfo  {journal} {\apj}\ }\textbf {\bibinfo {volume} {769}},\ \bibinfo {eid} {113} (\bibinfo {year} {2013})}\BibitemShut {NoStop}%
\bibitem [{\citenamefont {{Cappellaro}}\ \emph {et~al.}(2015)\citenamefont {{Cappellaro}}, \citenamefont {{Botticella}},\ and\ \citenamefont {{Pignata}}}]{Capellaro_2015IAUS..317..181}%
  \BibitemOpen
  \bibfield  {author} {\bibinfo {author} {\bibfnamefont {E.}~\bibnamefont {{Cappellaro}}}, \bibinfo {author} {\bibfnamefont {M.~T.}\ \bibnamefont {{Botticella}}},\ and\ \bibinfo {author} {\bibfnamefont {G.}~\bibnamefont {{Pignata}}},\ }\bibfield  {title} {\bibinfo {title} {{Supernova rates in the local Universe}},\ }\href {https://doi.org/10.1017/S1743921315000861} {\bibfield  {journal} {\bibinfo  {journal} {IAU Symposium}\ }\textbf {\bibinfo {volume} {317}},\ \bibinfo {pages} {181} (\bibinfo {year} {2015})}\BibitemShut {NoStop}%
\bibitem [{\citenamefont {{Heger}}\ \emph {et~al.}(2005)\citenamefont {{Heger}}, \citenamefont {{Woosley}},\ and\ \citenamefont {{Spruit}}}]{Heger_2005ApJ...626..350}%
  \BibitemOpen
  \bibfield  {author} {\bibinfo {author} {\bibfnamefont {A.}~\bibnamefont {{Heger}}}, \bibinfo {author} {\bibfnamefont {S.~E.}\ \bibnamefont {{Woosley}}},\ and\ \bibinfo {author} {\bibfnamefont {H.~C.}\ \bibnamefont {{Spruit}}},\ }\bibfield  {title} {\bibinfo {title} {{Presupernova Evolution of Differentially Rotating Massive Stars Including Magnetic Fields}},\ }\href {https://doi.org/10.1086/429868} {\bibfield  {journal} {\bibinfo  {journal} {\apj}\ }\textbf {\bibinfo {volume} {626}},\ \bibinfo {pages} {350} (\bibinfo {year} {2005})}\BibitemShut {NoStop}%
\end{thebibliography}%
\appendix

\section{Resolution study}
\label{app:resolution}

Two additional high-resolution simulations were performed to assess the dependence of the resonance on spatial resolution. We find that the resonance phenomenon persists, but the associated \gls{GW} flares become weaker and shorter-lived at higher resolution. This behaviour can be explained by small shifts in the $f$-mode frequency that prevent sustained overlap with the epicyclic frequency, highlighting the strong sensitivity of the resonance to resolution-dependent frequency variations.

All three simulations span the same domain: a radial extent from $\unit[0]{cm}$ to $\unit[10^{10}]{cm}$, and an azimuthal angle range of $[0,\,\pi]$. Model \texttt{IR} uses an innermost grid spacing of $\unit[400]{m}$, with $n_r=480$ radial zones and $n_\theta=128$ azimuthal zones. Model \texttt{IR-2} refines the innermost spacing to $\unit[300]{m}$, with $n_r\times n_\theta=602\times 256$, while Model \texttt{IR-3} maintains $\unit[300]{m}$ spacing but uses $n_r\times n_\theta=800\times 512$.

\begin{figure}[t]
\includegraphics[width=0.5\textwidth]{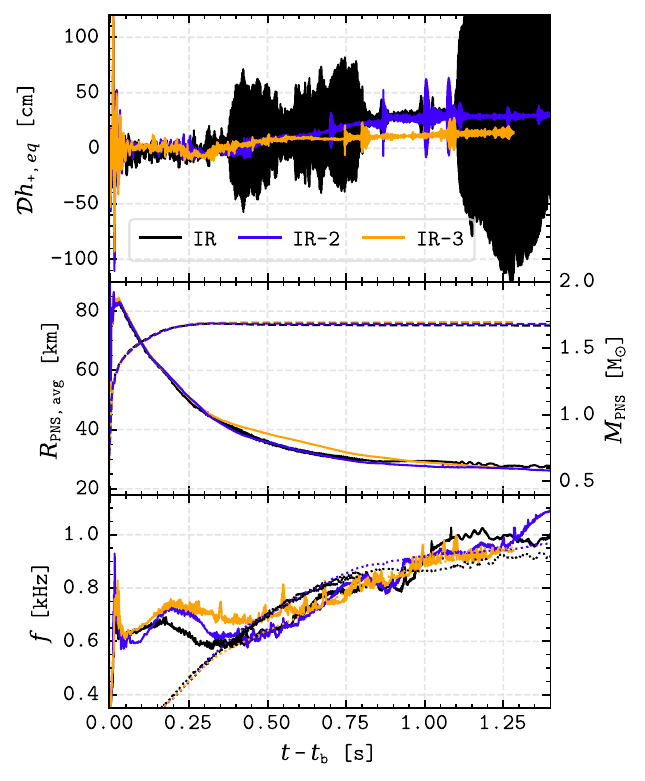} 
\caption{
    \label{fig:PNS_res} Top panel: \gls{GW} strains emitted by a source at distance $\mathcal{D}$ for models \texttt{IR} (Black), \texttt{IR-2} (blue), and \texttt{IR-3} (orange). Middle panel: evolution of \gls{PNS} radius (solid lines) and mass (dashed lines). Bottom panel: epicyclic frequency evolution (solid lines) and $f$-mode, computed using \cite{TorresForne19} (dotted).
}
\end{figure}

 Models \texttt{IR-2} and \texttt{IR-3} exhibit short flares of enhanced amplitude, lasting a few tens of milliseconds and reaching amplitudes of several tens of centimetres, but both the duration and amplitude are reduced compared to \texttt{IR} (Figure~\ref{fig:PNS_res} top panel). These flares are more similar in duration and amplitude to models \texttt{R3} and \texttt{R4}, in which the rotation has been altered (see next section).
The bottom panels display the spectrograms. In higher-resolution runs the $f$-mode frequency, indicated by  the darker band of increasing frequency, is shifted upward by $\sim\unit[100]{Hz}$, so the fundamental epicyclic frequency (solid blue lines) and its maximum (blue shading) do not remain aligned with the $f$-mode long enough to drive a strong resonance, producing only short, weak flares.

Consistency across resolutions is further confirmed in the bottom two panels of Figure~\ref{fig:PNS_res}. \gls{PNS} quantities for \texttt{IR} and \texttt{IR-2} agree within 5\%. Model \texttt{IR-3} shows a slightly larger and more oblate \gls{PNS}, due to its faster rotation. 
The epicyclic frequencies agree within $\unit[100]{Hz}$ overall, but in the first $\unit[0.3]{s}$ post-bounce, when the initial resonant bursts appear in \texttt{IR}, the higher-resolution models yield slightly higher values. This may delay the crossing with the $f$-mode beyond the computed times or happen in shorter periods, thereby damping the resonance.

As in many other resonant phenomena in physics (e.g., a driven, slightly damped pendulum or the propagation of seismic waves with anelastic attenuation), damping implies that a higher quality factor $Q$ results in a narrower frequency range over which true resonance occurs. The resolution study presented here suggests that the resonance observed has a relatively high $Q$, so even small detuning between the epicyclic frequency and the \gls{PNS} $f$-mode frequency significantly reduces the resonant amplitude. Consequently, time-dependent factors that slightly alter the conditions governing either the $f$-mode or the epicyclic frequency can induce partial quenching or intermittency in the resonant response.

This argument explains how changes in resolution—leading to the mild dynamical differences discussed in previous paragraphs—can cause a detuning between the \( ^2f \) and \( f_{\rm epi} \) modes and, accordingly, a significant reduction in the \gls{GW} amplitude. This result should not be interpreted as a lack of physical relevance of the identified resonance. Rather, it suggests that very slight modifications to the initial conditions—specifically, to the central nuclear rotational rate \( \Omega_{\rm c} \)—can trigger large-amplitude \glspl{GW} in these models (as observed for \( \Omega_{\rm c} = 1\,\mathrm{rad/s} \)). The substantial computational cost of higher-resolution models prevents a more detailed exploration to confirm this point. However, this interpretation is further supported by additional simulations based on different presupernova stellar models (not shown in this work) that exhibit similar resonant behaviour.

We also attribute to the high $Q$ the partial quenching of the resonance observed between the two main flares in model \texttt{IR}. The low-activity period coincides with a phase during which extended equatorial inflows strike the \gls{PNS} surface. Most of the high–specific-angular-momentum material impacting the \gls{PNS} is not assimilated (as indicated by the absence of a clear increase in \gls{PNS} mass); instead, it rebounds and is eventually expelled through the polar regions into the collimated supernova ejecta. We tentatively identify the random variations induced by these inflows on the \gls{PNS} as a plausible source of detuning between the $f$-mode and epicyclic frequencies.

In summary, the resonance is robust to changes in resolution, but its \gls{GW} signature depends sensitively on the relative values of the $f$-mode and epicyclic frequencies. 
Further high-resolution simulations are required to quantify the enhancement produced by this resonance and to constrain the parameter space in which it arises.

\section{Rotation dependence}
\label{app:rotation}

Two additional simulations exploring the rotation parameter space were carried out to test how the resonance depends on the initial rotation rate. These correspond to model \texttt{R3}, initialized with three times the rotation rate of the progenitor, and model \texttt{R4}, with a fourfold increase. These models are configured with nuclear rotational rates near the value at which we observed resonant behaviour, specifically $\Omega_{\rm c} = 0.87$, 1, and $1.16\,\mathrm{rad/s}$. The results show that the resonance is highly sensitive to rotation: increasing or decreasing the initial rotation changes both the timing and the character of the \gls{GW} signal.
\begin{figure}[hbt!]
\includegraphics[width=0.46\textwidth]{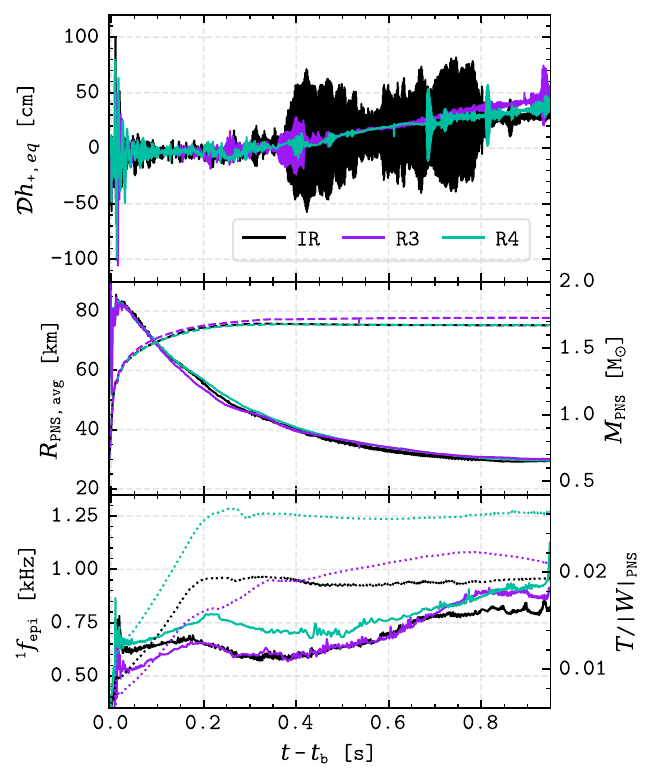} 
\caption{
    \label{fig:PNS_rot} Same as Figure~\ref{fig:PNS_res} but for models \texttt{IR}, \texttt{R3}, and \texttt{R4}, in which the $f$-mode has been replaced with $T/|W|$.
}
\end{figure}

The top panel of Figure~\ref{fig:PNS_rot} compares the \gls{GW} strains of models \texttt{R3}, \texttt{R4}, and \texttt{IR}. The reference model \texttt{IR} exhibits a continuous, sustained resonance over $\unit[0.4{-}0.8]{s}$. Model \texttt{R3}, instead, shows a strong but short-lived enhancement lasting about $\unit[100]{ms}$, beginning at $\unit[0.35]{s}$, followed by a quiescent phase until $\unit[0.9]{s}$. Model \texttt{R4} produces yet another pattern: a sequence of intermittent flares of elevated \gls{GW} amplitude, each lasting only a few tens of milliseconds. Together, these comparisons demonstrate that the resonance peaks for a specific rotation, while  weakens or fragments for increased or decreased spin.

The middle panel of Figure~\ref{fig:PNS_rot} shows the \gls{PNS} mass (dashed lines) and radius (solid lines). The three models produce very similar \gls{PNS} structures, with only a small mass offset: model \texttt{R3} is more massive by about $\sim\unit[0.05]{M_\odot}$, while the radii remain nearly identical.

The bottom panel of Figure~\ref{fig:PNS_rot} compares the epicyclic frequency (solid lines) and the ratio of rotational to gravitational energy, $T/|W|$ (dotted lines). In the slower-rotating models (\texttt{IR} and \texttt{R3}), the epicyclic frequencies are nearly identical between $\unit[0.2{-}0.6]{s}$. However, in model \texttt{R3}, $T/|W|$ gradually increases and eventually surpasses that of \texttt{IR}, even though \texttt{IR} began with faster rotation. After resonance is triggered, the value of $T/|W|$ in \texttt{IR} stabilizes, while it continues to rise in \texttt{R3}. In the fastest-rotating case, \texttt{R4}, both the epicyclic frequency and $T/|W|$ are consistently higher by about $\sim\unit[100]{Hz}$ and $0.005$, respectively. These differences indicate that changes in the \gls{PNS} shape and rotation rate can shift the frequency of the $f$-mode, altering the conditions required for the epicyclic frequency to maintain resonance. As a result, sustained enhancements of the \gls{GW} signal become harder to achieve at higher rotation rates.

The sensitivity to the rotational rate must be interpreted with care. Changes in numerical resolution propitiate a slight detuning between the $f$-mode and epicyclic frequencies, driven by mild differences in the dynamics. We suspect that resolution-dependent effects may lead to conditions in which the resonance is amplified for different initial values of $\Omega_{\rm c}$. The substantial computational resources required to verify this hypothesis compel us to defer a detailed study to future work.

In summary, the resonance is found within a relatively narrow range around $\Omega_{\rm c} \approx 1\,\mathrm{rad/s}$ under the idealized conditions of our models, where rotation is imposed on an initially non-rotating stellar structure. Combined with the uncertainties in the actual rotational profiles predicted by stellar evolution models, this makes it very difficult to estimate the fraction of massive, rotating stars in which conditions are favourable for the development of the resonance. Such conditions may occur in, say, $\sim 10\%$ of the tail of faster rotators (i.e., with equatorial projected rotational speeds $v \sin i > 150\,\rm km\, s^{-1}$; \cite{Holgado22}). This implies that only about $1\%$ of all massive stars could host favourable conditions for the resonance discussed in this work.

\end{document}